\documentclass[final]{aipproc}
\usepackage{epsfig,pstricks,graphicx}
\usepackage{amsmath,amssymb}

\layoutstyle{6x9}

\def\DJo{$\;$\kern-.4em \hbox{D\kern-.8em\raise.15ex\hbox{--}\kern.35em okovi\'c}}
\def\CC{{\rm\kern.24em \vrule width.04em height1.46ex depth-.07ex
\kern-.30em C}}
\def\RR{{\rm
         \vrule width.04em height1.58ex depth-.0ex
         \kern-.04em R}}
\def\P{{\rm I\kern-.25em P}}
\def\pR{{\mathcal R}}
\def\id{{\rm 1\kern-.22em l}}
\def\i{{\rm i}\,}

\def\trace{{\rm tr}\;}
\def\pH{{\mathcal H}}
\def\sig{{\sigma}}
\let\la=\langle  \let\ra=\rangle
\def\vf{{\varphi}}
\newtheorem{proposition}{Proposition}[section]

\newtheorem{psatz}{Satz}[section]
\newtheorem{pdef}{Definition}[section]
\newtheorem{conjecture}{Vermutung}[section]

\newcommand{\Matrix}[2]{\left( \begin{array}{#1} #2 \end{array}
  \right)}
\newcommand{\fat}[1]{\mbox{\boldmath $ #1 $\unboldmath}}
\newcommand{\beq}{\begin{equation}}
\newcommand{\beqa}{\begin{eqnarray}}
\newcommand{\nbeqa}{\begin{eqnarray*}}
\newcommand{\eeq}{\end{equation}}
\newcommand{\eeqa}{\end{eqnarray}}
\newcommand{\neeqa}{\end{eqnarray*}}
\newcommand{\bra}[1]{\left\langle #1 \right |}
\newcommand{\ket}[1]{\left | #1 \right\rangle}

\newcommand{\expect}[1]{\left\langle #1 \right\rangle}

\newcommand{\bigfrac}[2]{\mbox {${\displaystyle \frac{ #1 }{ #2 }}$}}
\newenvironment{eqblock}[2]{\beq\label{#2}\begin{array}{#1}}{\end{array}
                                \eeq}
\newenvironment{neqblock}[1]{\[\begin{array}{#1}}{\end{array}\]}
\newcommand{\beqb}{\begin{eqblock}}
\newcommand{\eeqb}{\end{eqblock}} 
\newcommand{\nbeqb}{\begin{neqblock}}
\newcommand{\neeqb}{\end{neqblock}} 
\newtheorem{theorem}{Theorem}[section]

\newcommand{\Determinant}[2]{\left| \begin{array}{#1} #2 \end{array}
  \right|}

\begin{document}

\title{On polynomial invariants of several qubits}
\author{D. \v{Z}. \DJo}{
  address={Department of Pure Mathematics, University of Waterloo, Waterloo, Ontario, N2L 3G1, Canada, email: djokovic@uwaterloo.ca}}
\author{A. Osterloh}{
address={Institut f\"ur Theoretische Physik, Leibniz Universit\"at Hannover, D-30167 Hannover, Germany, email: andreas.osterloh@itp.uni-hannover.de}}

\begin{abstract}
It is a recent observation that entanglement classification for qubits
is closely related to local $SL(2,\CC)$-invariants including the
invariance under qubit permutations~\cite{Duer00,OS04,Dokovic},
which has been termed $SL^*$ invariance.
In order to single out the $SL^*$ invariants, we analyze the $SL(2,\CC)$-invariants
of four resp. five qubits and decompose them into irreducible modules
for the symmetric group $S_4$ resp. $S_5$ of qubit permutations.
A classifying set of measures of genuine multipartite entanglement is given
by the ideal of the algebra of $SL^*$-invariants vanishing on arbitrary product states.
We find that low degree homogeneous components of this ideal can be
constructed in full by using the approach introduced in Refs.~\cite{OS04,OS05}.
Our analysis highlights an intimate connection between this latter procedure
and the standard methods to create invariants, such as the $\Omega$-process
~\cite{Luque05}. As the degrees of invariants increase,
the alternative method proves to be particularly efficient.
\end{abstract}
\classification{03.67.Mn,02.10.Hh}
\keywords{Entanglement -- multipartite -- polynomial invariants -- $\Omega$-process}

\maketitle

\section{Introduction}
The quantification and classification of multipartite entanglement 
takes an important part in quantum information theory and is subject
to a lively discussion in the recent literature.
Many different sets of measures of entanglement have been proposed for 
introducing some order and insight into the Hilbert space of 
multipartite systems.
An important part of the discussion addresses the underlying 
invariance group the measures have to have. Local unitary invariance
is clearly a minimal requirement but must be extended
to invariance under local special linear transformations, 
when generalized local operations are admitted.
For qubit systems this local invariance group is the $SL(2,\CC)$.
The invariance group of $q$ qubits is then given by 
$SL_q:=SL(2,\CC)^{\otimes q}$, where we will even omit the index
wherever it doesn't create confusion \footnote{That is: where either it 
is clear what number of qubits we are talking about, or in generic statements
applying to arbitrary number of qubits. We deviate from the
standard definition $SL_q:=SL(q,\CC)$; since we deal exclusively 
with qubits throughout the paper, this should not cause any confusion.}.
Interestingly, the demand of invariance under $SL$ operations
on the measures of pure state entanglement readily implies that
the induced measure on mixed states is an entanglement monotone
when extended through its convex roof~\cite{MONOTONES,VerstraeteDM03}.
The requirement of $SL$-invariance is restrictive enough to
even single out a distinguished class of genuine multipartite entangled states:
the nonzero SLOCC classes are made of all those states that do not vanish 
after infinitely many $SL$ operations (the
{\em local filtering operations} of e.g. Ref.~\cite{VerstraeteDM03}).
Each such nonzero Stochastic Local Operation and Classical Communication (SLOCC) 
class has a representative
which can be considered as maximally entangled state within that class.
Interestingly, for three qubits a maximally three-tangled state
with no concurrence exists, and for four qubits there are three 
four-tangled states with neither three-tangle nor concurrence. 
It is an open question whether such representatives 
carrying only the genuine multipartite entanglement classes will exist
in general.
  
It is worth emphasizing at this point that any function
of the pure state coefficients that is invariant under
$SL$ transformations will remain unchanged by such local filtering 
operations. Consequently, the complementary 
{\em zero SLOCC class} exists that contains all states for which
all these invariant functions are zero.
A prime example for a representative of the zero SLOCC class
is the multiqubit W state
$\ket{W}=\sum_i \alpha_i\ket{i}$ with $\ket{i}=\ket{0\dots 1\dots 0}$
being a state with all zeros but a single 1 placed at site number $i$
(or straightforward generalizations of it to higher local dimension).
Notwithstanding its globally distributed entanglement of pairs
we therefore would not call it genuinely multipartite entangled.

$SL$-invariance has been intensely studied for three qubit
systems in Refs.~\cite{Duer00,Wallach} and for four qubit systems in 
Refs.~\cite{Dokovic,Luque02}, and geometric aspects of such invariants
have been highlighted in Refs.~\cite{Levay04,Levay05,Levay06}. 
Preliminary results for five qubits
have been presented recently~\cite{Luque05}.
Independent of these approaches, a method based on local $SL(2,\CC)$-invariant
operators has been suggested with emphasis on permutation
invariance of the global entanglement measure~\cite{OS04,OS05}. 
Permutation invariance has been highlighted as a demand 
on global entanglement measures already in Ref.~\cite{Coffman00} and later in 
Ref.~\cite{Dokovic}, where the semidirect product of $SL_q$ and
the symmetric group $S_q$ of qubit permutations has been termed
$SL^*_q:=SL(2,\CC)^{\otimes q}\times S_q$, which we will 
abbreviate as $SL^*$. In addition, Refs.~\cite{OS04,OS05} focus 
on those invariants that vanish on all product states.
These form an ideal in the ring of $SL^*$-invariants and are important 
for distinguishing
genuine multipartite entangled states from tensor products of entangled states
such as $\ket{GHZ}\otimes\ket{GHZ}$. It lies within the nonzero SLOCC class of 
$6$ qubit entangled states but is not genuinely $6$ qubit entangled.

In this work we will use both local invariant operators as
proposed in Refs.~\cite{OS04,OS05} and the Cayley $\Omega$-process to construct
polynomial invariants. First we compare both approaches for the known complete set of
invariants of four qubits~\cite{Luque02} and those for five qubits 
up to degree $6$ (see Ref.~\cite{Luque05}). 
For these known cases we follow the notation of Refs.~\cite{Luque02,Luque05}
and express the invariants presented there
in terms of combs and filters from Refs.~\cite{OS04,OS05}.
Then we go considerably ahead up to degree $12$ with an outlook to degrees $14$ and $16$.

The manuscript is organized as follows: the next Section reviews the 
approach to $SL$-invariants using local invariant operators and fixes the notation.
Section \ref{fourbits} summarizes the main results for 4-qubit invariants from 
Ref.~\cite{Luque02} and establishes the correspondence between these invariants 
and those from Ref.~\cite{OS04}.
Section \ref{fivebits} briefly revisits the known polynomial invariants (up to degree $6$)
and then gives a new and complete characterization for the space of
five-qubit $SL$-invariants for degrees $8,10$ and $12$. We also make a few remarks
regarding the invariants of degree $14$ and $16$. In Section~\ref{SLstarseries}
we determine the Hilbert series for $SL^*$-invariants which confirms the findings of the preceding Sections.
After presenting an interesting connection between the 
Cayley $\Omega$-process and local invariant
operators (combs) in Section \ref{altomega} we draw our conclusions
in Section \ref{concls}.
The appendix provides a detailed discussion of the concepts and notations
of the comb based approach in Refs.~\cite{OS04,OS05} together with 
a prescription for their evaluation. 

\section{Local invariant operators and notation}\label{notation}

Before we start with our analysis, we give a brief summary of the approach 
using local invariant operators. For a more detailed description 
of this approach see the appendix~\ref{combs}.

We will refer to multiple copies of a given quantum state
simply as {\em copies} in what follows.
The Hilbert space for $m$ copies of a quantum state $\psi$ of $q$ qubits
can be written as
$$
(\pH_{11} \otimes \pH_{12} \otimes \cdots \otimes \pH_{1q} ) 
\bullet ( \pH_{21} \otimes \pH_{22} \otimes \cdots \otimes \pH_{2q} )\bullet 
\cdots 
\bullet ( \pH_{m1} \otimes \pH_{m2} \otimes \cdots \otimes \pH_{mq} ) ,
$$
where $\pH_{i1} \otimes \pH_{i2} \otimes \cdots \otimes \pH_{iq}$
is the Hilbert space for the $i$-th copy of $\psi$, $\pH_{ij}$ is the Hilbert space 
for the $j$-th qubit in this copy, and $\bullet$ is used as the tensor 
product sign between the Hilbert spaces of different copies of $\psi$.

We will often use the notion of an {\em expectation value} of an 
operator $\hat{O}$, which for a pure state $\ket{\psi}$ is defined as 
$\bra{\psi}\hat{O}\ket{\psi}$.
A qubit {\em comb} has been defined in Refs.~\cite{OS04,OS05} as an antilinear 
operator acting on a single or multiple copy of a pure single qubit state ($q=1$) 
which has zero expectation value for all such states.
We point out that combs are $SL(2,\CC)$-invariant operators.
Two independent combs $\sigma_2 \mathfrak{C}$ and 
$\sigma_\mu\mathfrak{C}\bullet\sigma^\mu\mathfrak{C}$
have been identified in terms of the Pauli matrices
\beq\label{Pauli}
\sigma_0:=\Matrix{cc}{1&0\\0&1}\; ,\quad
\sigma_1:=\Matrix{cc}{0&1\\1&0}\; ,\quad
\sigma_2:=\Matrix{cc}{0&-i\\i&0}\; ,\quad
\sigma_3:=\Matrix{cc}{1&0\\0&-1}\; ,
\eeq
where $\mathfrak{C}$ is the complex 
conjugation in the eigenbasis of $\sigma_3$, and the contraction
is defined via the pseudo-metric $G_{\mu\nu}:=\delta_{\mu\nu} g_\mu$ as
\beqa\label{contract}
\sigma_\mu\bullet\sigma^\mu&:=&\sum_{\mu=0}^3 g_\mu \sigma_\mu\bullet\sigma_\mu\\
(g_0,g_1,g_2,g_3)&:=&(-1,1,0,1)\label{metric}
\eeqa
Being $SL$-invariant, both combs admit
the construction of antilinear operators acting on multiple copies
of pure multiqubit states that are $SL$-invariant.
Polynomial invariants are then constructed from multiqubit 
operators obtained from combs as their 
antilinear expectation values for a general multiqubit pure state. 
These invariants are homogeneous polynomials in the basis coefficients of 
the state $\ket{\psi}$ (see the appendix for more details).
We will use double brackets to denote the expectation value of
an antilinear operator and we often omit the tensor product sign $\otimes$ :
\beqa\label{singlecopy}
(\!(\sigma_2\sigma_2)\!)&:=&(\!(\sigma_2\otimes\sigma_2)\!):=
\bra{\psi^*}\sigma_2\otimes\sigma_2\ket{\psi}\\
(\!(\sigma_\mu\sigma_\nu\dots\bullet\sigma^\mu\sigma^\nu\dots)\!)&:=&
\bra{\psi^*}\bullet\bra{\psi^*}(\sigma_\mu\otimes\sigma_\nu\otimes\dots)\bullet
(\sigma^\mu\otimes\sigma^\nu\otimes\dots)\ket{\psi}\bullet\ket{\psi}\;
\label{doubleparenthesis} .
\eeqa
Here, $\ket{\psi}$ is a pure state of $q\geq 2$ qubits (expressible as a vector
in $(\CC^2)^{\otimes q}$). 
The double bracket expressions in Eq.~\eqref{doubleparenthesis} can be 
evaluated by first calculating the (antilinear) expectation values
for the two copies of $\ket{\psi}$ and performing afterwards the contractions 
with the pseudometric.
In \eqref{singlecopy} and \eqref{doubleparenthesis} we have shown 
expressions for one and two copies of the state only. 
The extension to more than two copies 
(and hence higher degree of the invariant) is defined analogously.
A measure of entanglement is then defined as the absolute value of such an 
invariant, e.g. $C=|(\!(\sigma_2\sigma_2)\!)|$ 
is the pure state concurrence~\cite{Hill97}.

By a {\em product state} we mean a state that can be written 
as a tensor product on some bipartition of the system. 
We slightly relax the use of the term ``filter'' 
as compared with Ref.~\cite{OS04},
where also permutation invariance was included.
We will call a {\em filter} an invariant which vanishes on all
product states and reserve the term {\em ${\rm SL}^*$-filter} for 
those which are also invariant under qubit permutations.
The algebra of complex holomorphic polynomial
$SL$-invariants resp. $SL^*$-invariants of $q$
qubits will be denoted by
${\rm Inv}^{SL}$ resp. ${\rm Inv}^{SL^*}$.
The number of qubits, $q$, will be clear from
the context. 
The filters form an ideal ${\cal I}^{SL}_0$ of the algebra ${\rm Inv}^{SL}$. 
The ${SL^*}$-filters form an ideal
${\cal I}^{SL^*}_0:={\cal I}^{SL}_0 \cap {\rm Inv}^{SL^*}$ 
of ${\rm Inv}^{SL^*}$.
The subspace of ${\rm Inv}^{SL}$, ${\rm Inv}^{SL^*}$, etc.
consisting of homogeneous invariants of
degree $d$ will be denoted by adding the degree as an index. E.g. 
${\rm Inv}_d^{SL}$ and ${\rm Inv}_d^{SL^*}$. 

Let us point out four important facts. First, ${\rm Inv}^{SL}_d=0$ 
whenever $d$ is odd. Second, ${\rm Inv}^{SL}_2=0$ if $q$ is odd
and has dimension $1$ if $q$ is even. 
These facts are special cases of more general results proved in 
\cite[Prop. 11.1 and Cor. 11.2]{Brylinski02b}.
Third, the dimension of ${\rm Inv}^{SL}_4$ is equal to $(2^{q-1}+(-1)^q)/3$.
Fourth, the dimension of ${\rm Inv}^{SL^*}_4$ is equal to 
$\lfloor \frac{q+5}{6} \rfloor$, where $\lfloor t\rfloor$ 
denotes the Gauss parenthesis, i.e. the largest integer $n\leq t$.
These two results are proved in \cite[Cor. 11.4 and Prop. 11.3]{Brylinski02b}.

We will also use the notion of a {\em relative} $SL^*$-{\em invariant} 
for an $SL$-invariant that is fixed up to a sign under all qubit permutations. 
Antisymmetric relative invariants will be termed
{\em odd relative invariants} or $SL^*_-$-invariants.
In this context we will sometimes need to either symmetrize or
antisymmetrize a given invariant for obtaining the corresponding
$SL^*$ and $SL^*_-$ invariants, respectively. For an operator $\hat{O}$, whose 
dependence on qubit permutations is indicated by the permutation operator 
as an index, we use the definitions
\nbeqa
\expect{\hat{O}_\id}_s&:=&\frac{1}{q!}\sum_{\pi\in S_q}\hat{O}_\pi , \\
\expect{\hat{O}_\id}_a&:=&\frac{1}{q!}\sum_{\pi\in S_q}{\rm sign}\,\pi
\;\hat{O}_\pi \ .
\neeqa
An $S_q$ orbit of an invariant $\hat{I}$ will be denoted by $S_q\circ\hat{I}$,
and $\pi_{ij}$ will denote the permutation operator that exchanges qubit 
numbers $i$ and $j$.
Furthermore we will say that an invariant is {\em generically of degree $d$},
if it is not expressible as a polynomial in invariants of lower degrees. 
To simplify the notation, we set $V_d:={\rm Inv}_d^{SL}$ and denote by $U_d$
the subspace of $V_d$ spanned by the products of homogeneous lower degree 
invariants, i.e. by $V_{s} V_{d-s}$ for $s=1,\dots ,d-1$. 
For the decomposition of the space $V_d$ (or $U_d$)
into simple modules $X_i$ of the symmetric group $S_q$
we use the notation of Ref.~\cite{JamesKerber}.

\section{$SL$ and $SL^*$-invariants for four qubits}
\label{fourbits}

The Hilbert series for $SL_4$-invariants is~\cite{Luque02}
\nbeqa
h(t)&=&\frac{1}{(1-t^2)(1-t^4)^2(1-t^6)}\\
&=& 1+t^2+3t^4+4t^6+7t^8+9t^{10}+14t^{12}+17 t^{14} +24 t^{16} +29 t^{18}\\ 
&&+ \dots 
\neeqa
From theorem 4.2 of Ref.~\cite{Dokovic} we obtain immediately
the Hilbert series for $SL^*_4$-invariants
\nbeqa
h_{SL^*}(t)&=&\frac{1}{(1-t^2)(1-t^6)(1-t^8)(1-t^{12})}\\
&=& 1+t^2+t^4+2t^6+3t^8+3t^{10}+5t^{12}+6 t^{14} +7 t^{16} +9 t^{18}\\
&& + \dots 
\neeqa
We deduce that the algebra ${\rm Inv}^{SL_4}$ is a polynomial algebra with 
generators of degree $2$, $4$, $4$ and $6$ and, similarly, that 
${\rm Inv}^{SL^*_4}$ is a polynomial algebra with 
generators of degree $2$, $6$, $8$ and $12$.

Furtheremore, a complete set of invariants~\cite{Luque02} and covariants~\cite{BriandLT03} is known.
With the focus of finding measures for genuine multipartite entanglement,
three independent filter invariants have been constructed in Ref.~\cite{OS04} 
\nbeqa
{\cal F}^{(4)}_1  &=& 
                (\!(\sigma_\mu\sigma_\nu\sigma_2\sigma_2
                  \bullet \sigma^\mu\sigma_2\sigma_\lambda\sigma_2\bullet 
                \sigma_2\sigma^\nu\sigma^\lambda\sigma_2)\!) \\
{\cal F}^{(4)}_2 &=&
                (\!(\sigma_\mu\sigma_\nu\sigma_2\sigma_2
                  \bullet \sigma^\mu\sigma_2\sigma_\lambda\sigma_2\bullet 
   \\
   &&\qquad\qquad       \sigma_2\sigma^\nu\sigma_2\sigma_\tau 
            \bullet
                \sigma_2\sigma_2\sigma^\lambda\sigma^\tau)\!)\\
{\cal F}^{(4)}_3 &=&\bigfrac{1}{2}
                (\!(\sigma_\mu\sigma_\nu\sigma_2\sigma_2
                  \bullet \sigma^\mu\sigma^\nu\sigma_2\sigma_2
\bullet \sigma_\rho\sigma_2\sigma_\tau\sigma_2 \bullet
         \\ &&\qquad  
          \sigma^\rho\sigma_2\sigma^\tau\sigma_2
 \bullet \sigma_\kappa\sigma_2\sigma_2\sigma_\lambda \bullet
                \sigma^\kappa\sigma_2\sigma_2\sigma^\lambda)\!)
 \ \ . 
\neeqa
We will use this Section to work out interrelations between the two approaches.

\subsection{Degree $2$}
As mentioned in the end of Section~\ref{notation}, 
this smallest possible degree appears only for an even number of qubits
$q$ and the corresponding space $V_2$ is one-dimensional with 
the $q$-tangle of Wong and Christensen~\cite{Wong00} as generator.
Here, for $q=4$, this generator is the $4$-tangle and has 
been termed $H$ in Ref.~\cite{Luque02}
$$
H(\psi)=
\frac{1}{2}(\!(\sigma_2\sigma_2\sigma_2\sigma_2)\!)=:
\frac{1}{2} {\cal C}^{(4)}_{2} \; .
$$
It does not vanish on tensor products of $2$-qubit entangled states and so it 
is not a filter.

Summarizing, we have that 
\nbeqa
{\rm Inv}_{2}^{SL^*}&=&{\rm Inv}_{2}^{SL}={\rm span}\{H\}\\
{\cal I}_{0;2}^{SL}&=& 0 \ .
\neeqa

\subsection{Degree $4$}
Besides the one-dimensional space $U_4$ spanned by $H^2$, 
there exist three new invariants of degree $4$,
namely $L$, $M$, and $N$ subject to the relation $L+M+N=0$.
Expressed in terms of the coefficients of the wave function
$$
\ket{\psi}=\sum_{i,j,k,l=0}^1\psi_{ijkl}\ket{ijkl}=:\sum_{n=0}^{2^4-1} a_n \ket{n},
$$
with the identification $\ket{ijkl}\equiv\ket{i+2j+4k+8l}$,
they are given by the determinants
\beq
L=\Determinant{cccc}{a_0 & a_4 & a_8 & a_{12}\\
                     a_1 & a_5 & a_9 & a_{13}\\
                     a_2 & a_6 & a_{10} & a_{14}\\
                     a_3 & a_7 & a_{11} & a_{15}}, \;
M=\Determinant{cccc}{a_0 & a_8 & a_2 & a_{10}\\
                       a_1 & a_9 & a_3 & a_{11}\\
                       a_4 & a_{12} & a_{6} & a_{14}\\
                       a_5 & a_{13} & a_{7} & a_{15}}, \;
N=\Determinant{cccc}{a_0 & a_1 & a_8 & a_{9}\\
                       a_2 & a_3 & a_{10} & a_{11}\\
                       a_4 & a_5 & a_{12} & a_{13}\\
                       a_6 & a_7 & a_{14} & a_{15}}.
\eeq
They can be expressed in terms of the following invariants obtained from 
local invariant operators 
\nbeqa
{\cal C}^{(4)}_{4;(1,2)} &:=& (\!(\sigma_\mu\sigma_\nu\sigma_2\sigma_2
                              \bullet \sigma^\mu\sigma^\nu\sigma_2\sigma_2)\!), \\
{\cal C}^{(4)}_{4;(1,3)} &:=& (\!(\sigma_\mu\sigma_2\sigma_\nu\sigma_2
                              \bullet \sigma^\mu\sigma_2\sigma^\nu\sigma_2)\!), \\
{\cal C}^{(4)}_{4;(1,4)} &:=& (\!(\sigma_\mu\sigma_2\sigma_2\sigma_\nu
                              \bullet \sigma^\mu\sigma_2\sigma_2\sigma^\nu)\!).
\neeqa
Indeed we have
$$
L=\frac{1}{48}\left[{\cal C}^{(4)}_{4;(1,3)}-{\cal C}^{(4)}_{4;(1,4)}\right], \quad
M=\frac{1}{48}\left[{\cal C}^{(4)}_{4;(1,4)}-{\cal C}^{(4)}_{4;(1,2)}\right], \quad
N=\frac{1}{48}\left[{\cal C}^{(4)}_{4;(1,2)}-{\cal C}^{(4)}_{4;(1,3)}\right]
$$
and
$$
H^2=\frac{1}{12}\left[
{\cal C}^{(4)}_{4;(1,2)}+{\cal C}^{(4)}_{4;(1,3)}+{\cal C}^{(4)}_{4;(1,4)}\right].
$$
For analoguously defined invariants ${\cal C}^{(4)}_{4;(3,4)}$, 
${\cal C}^{(4)}_{4;(2,4)}$ and ${\cal C}^{(4)}_{4;(2,3)}$ we have the identities 
$$ {\cal C}^{(4)}_{4;(1,2)}\equiv{\cal C}^{(4)}_{4;(3,4)},\quad
{\cal C}^{(4)}_{4;(1,3)}\equiv{\cal C}^{(4)}_{4;(2,4)},\quad
{\cal C}^{(4)}_{4;(1,4)}\equiv{\cal C}^{(4)}_{4;(2,3)}.$$

It is interesting to mention at this point that further identities 
appear besides those stated above. Examples are
\beq\label{mmmmId}
(\!(\sigma_\mu \sigma_\nu \sigma_\lambda \sigma_\tau
\bullet \sigma^\mu \sigma^\nu \sigma^\lambda \sigma^\tau)\!)
= 36 H^2\; , 
\eeq
and the identity for the three-tangle in~\cite{OS04}. 
We will report on such identities also for five qubit invariants.
They suggest that double contractions 
$(\sigma_\mu\otimes\sigma_\nu)\bullet(\sigma^\mu\otimes\sigma^\nu)$ within
a pair of copies could be somehow removed. However, the non-trivial example 
$(\!(\sigma_2\sigma_2\sigma_2\sigma_2)\!)\neq(\!(\sigma_\mu\sigma_\nu\sigma_2\sigma_2)\!)
(\!(\sigma^\mu\sigma^\nu\sigma_2\sigma_2)\!)$ 
demonstrates that double contractions can not simply be removed.  
It would be worthwhile analyzing this curious observation in more detail and
a rigorous reduction scheme would be highly desirable.
The interrelation between the $\Omega$-process and the
comb approach in Section~\ref{altomega} singles out one origin for such 
identities. In particular it explains all the identities for degree 
$4$ invariants mentioned here.

Summarizing, we have
\nbeqa
{\rm Inv}_{4}^{SL}&=&{\rm span}\{{\cal C}^{(4)}_{4;(1,2)},
{\cal C}^{(4)}_{4;(1,3)},{\cal C}^{(4)}_{4;(1,4)}\}\\ 
{\rm Inv}_{4}^{SL^*}&=&{\rm span}\{H^2\}\subseteq U_4\\
{\cal I}_{0;4}^{SL}&=& 0\ .
\neeqa

\subsection{Degree $6$}
By invoking the Hilbert series, we deduce that $\dim U_6=3$. We claim
that $V_6$ is spanned by $U_6$ and the filter ${\cal F}^{(4)}_1$ from Ref.~\cite{OS04}.
As $\dim V_6=4$, it suffices to observe that ${\cal F}^{(4)}_1\notin U_6$.

Defining the $SL^*$-invariant $W:=D_{xy}+D_{xz}+D_{xt}$, 
the expressions for the $D_{uv}$ from Ref.~\cite{Luque02} give
\nbeqa
H(N-M)&=&3 D_{xy}- W, \\
H(L-N)&=&3 D_{xz}- W, \\
H(M-L)&=&3 D_{xt}- W.
\neeqa
For comparison with Ref.~\cite{Dokovic} the correspondence for the
invariants are $D_{xt}\rightarrow D$, $D_{xy}\rightarrow E$, 
$D_{xz}\rightarrow F$ and $W\rightarrow \Gamma$.

The subspace of $SL$-invariants of degree $6$ is spanned by 
$D_{xy}$, $D_{xz}$, $D_{xt}$, and $H^3$~\cite{Luque02} and we find that
$$
{\cal F}^{(4)}_1=8(4 W -  H^3).
$$
{}From these relations all invariants in this subspace are readily expressed
in terms of comb-based invariants.
It is worth noticing that ${\cal F}^{(4)}_1$ is a filter and that
it spans the subspace ${\cal I}^{SL}_{0;6}$.

Summarizing, we have
\nbeqa
{\rm Inv}_{6}^{SL}&=&{\rm span}\{H^3,H{\cal C}^{(4)}_{4;(1,2)},H{\cal C}^{(4)}_{4;(1,3)},{\cal F}^{(4)}_1\}\\ 
{\rm Inv}_{6}^{SL^*}&=&{\rm span}\{H^3,{\cal F}^{(4)}_1\} \\
{\cal I}_{0;6}^{SL}&=&{\rm span}\{{\cal F}^{(4)}_1\}\ .
\neeqa

All spaces
${\rm Inv}^{SL}_{d}$ with $d>6$ are built from generators of degree
$2$, $4$, and $6$. What we will focus on in the rest of this section
is to construct a complete set of generators for the ideal
${\cal I}^{SL^*}_0$.

\subsection{Degree $8$}
In this case we have $U_8=V_8={\rm Inv}^{SL}_8$ and $\dim V_8=7$.
By proposition \ref{IdGen}, which is proved below,
the degree $8$ component of ${\cal I}^{SL^*}_0$ is only two-dimensional.
It is spanned by $H{\cal F}^{(4)}_1$ and the symmetrized filter 
$\expect{{\cal F}^{(4)}_{2}}_s$. 
Defining $\Sigma:=L^2+M^2+N^2$ we find that
\nbeqa
{\cal F}^{(4)}_2 &=& 16 \left( H^4+4 H^2(M-L) -16 H D_{xt} -16 L M \right), \\ 
\expect{{\cal F}^{(4)}_{2}}_s&=&
\frac{16}{3} (8 \Sigma - H^4) - \frac{64}{3} H ( 4 W - H^3)\; .
\neeqa
The action of the symmetric group $S_4$ on the filter 
${\cal F}^{(4)}_{2}$ produces three independent filter invariants.
We have
\nbeqa
{\rm Inv}_{8}^{SL}&=&{\rm span}\{H^4,H^2{\cal C}^{(4)}_{4;(1,2)},
H^2{\cal C}^{(4)}_{4;(1,3)},H{\cal F}^{(4)}_1,S_4\circ {\cal F}^{(4)}_{2}\}\\ 
{\rm Inv}_{8}^{SL^*}&=&{\rm span}\{H^4,H{\cal F}^{(4)}_1,\expect{{\cal F}^{(4)}_{2}}_s\}\\
{\cal I}_{0;8}^{SL^*}&=&{\rm span}\{H{\cal F}^{(4)}_1,\expect{{\cal F}^{(4)}_{2}}_s\}\ .
\neeqa
We note that the orbit $S_4\circ {\cal F}^{(4)}_{2}$ in the first formula 
above can be replaced by the three invariants: 
$\left( {\cal C}^{(4)}_{4;(1,2)} \right)^2$,
$\left( {\cal C}^{(4)}_{4;(1,3)} \right)^2$ and
${\cal C}^{(4)}_{4;(1,2)}{\cal C}^{(4)}_{4;(1,3)}$.

\subsection{Degree $10$}
The degree $10$ homogeneous component ${\cal I}_{0;10}^{SL^*}$ is 
two-dimensional and is
spanned by $H^2{\cal F}^{(4)}_{1}$ and $H\expect{{\cal F}^{(4)}_{2}}_s$. 
The last missing ideal generator is obtained from degree $12$.

\subsection{Degree $12$ and beyond}
The $SL^*$-invariants of degree $12$ are to be built from
$H$, $W$, $\Sigma$ and $\Pi:=(L-M)(M-N)(N-L)$.
The filter ${\cal F}^{(4)}_3=\frac{1}{2} {\cal C}^{(4)}_{4;(1,2)}
{\cal C}^{(4)}_{4;(1,3)}{\cal C}^{(4)}_{4;(1,4)}$ is invariant
under qubit permutations, i.e. it is an $SL^*$-filter. We find that
$$
{\cal F}^{(4)}_3=-96 H^2 (8 \Sigma - H^4) - 64 (32 \Pi + H^6)\ .
$$

\begin{proposition} \label{IdGen}
The ideal ${\cal I}^{SL^*}_0$ is generated by the
invariants $4W-H^3$, $8\Sigma-H^4$, and $32\Pi+H^6$.
\end{proposition}

{\em Proof: }First, it is easy to check that these three invariants belong to 
${\cal I}^{SL^*}_0$. Next, let $f\in{\cal I}^{SL^*}_0$ be arbitrary. Note
that $f$ is a polynomial in the generators $H,W,\Sigma$ and $\Pi$ of 
the algebra ${\rm Inv}^{SL^*}$. Without any loss of generality we can assume 
that $f$ is homogeneous of degree $2d$.
The above three invariants can be used to eliminate
$W$, $\Sigma$ and $\Pi$ from $f$. Then the corresponding reduced element 
$f_0\in{\cal I}^{SL^*}_0$ is a homogeneous polynomial in $H$ only.
Consequently, $f_0=c H^d$ for some constant $c$.
In particular, $f_0$ must vanish on arbitrary product states.
Since $H$ however does not vanish on arbitrary product states,
this implies $c=0$ and completes the proof.

Equivalently, the same ideal is generated by 
the $SL^*$-filters ${\cal F}^{(4)}_1$, 
$\expect{{\cal F}^{(4)}_{2}}_s$ and ${\cal F}^{(4)}_3$ which are
functionally independent~\cite{OS04}. This follows immediately from
\nbeqa
4 W -  H^3&=&\frac{1}{8}{\cal F}^{(4)}_1 , \\
8 \Sigma - H^4&=&\frac{3}{16}\left[
        \expect{{\cal F}^{(4)}_{2}}_s+\frac{8}{3} H {\cal F}^{(4)}_1\right] , \\
32 \Pi + H^6&=&-\frac{1}{64}
        \left[{\cal F}^{(4)}_3 + 
        18 H^2\expect{{\cal F}^{(4)}_{2}}_s+48 H^3 {\cal F}^{(4)}_1\right].
\neeqa

As an important and often cited invariant,
we briefly consider the hyperdeterminant, {\bf Det}, of four qubits.
It has degree $24$ and is given by
$$
2^8 3^3 {\fat{\rm Det}} = (H^3 -4 W) A + (8\Sigma-H^4) B - 4 (32\Pi+H^6)^2 \; ,
$$
where
\nbeqa
A&=&5H^9+20W H^6-144\Sigma H^5+16(5 W^2-24\Pi) H^3\\
     && -960 W\Sigma H^2+1536\Sigma^2 H+192 W(3W^2+8\Pi)\; ,\\
B&=&H^8 -136\Sigma H^4 -384\Pi H^2 +256\Sigma^2\; .
\neeqa
This can be translated into an expression in terms of $H$ and the filters
${\cal F}^{(4)}_1$, $\expect{{\cal F}^{(4)}_{2}}_s$ and ${\cal F}^{(4)}_3$
in a straightforward manner.

The decomposition of ${\rm Inv}_d^{SL}$ for even $d$,
$2\le d\le 12$, into irreducible $S_4$-modules is given in table \ref{inv4}. 
Note that the $SL^*$ Hilbert series confirms the multiplicities of the 
trivial module $X_1$.

\begin{table}[h]
\caption{The space of polynomial invariants of degree $2$ up
to $12$ into irreducible $S_4$-modules.}\label{inv4}
\begin{tabular}{|c|l||c|l||c|l|}
\hline
degree &  & degree &  & degree &  \\
\hline
\hline
2 &  $X_1$ &  4 & $X_1 + X_3$ &  6 & $2X_1 + X_3$ \\
\hline
     8 & $3X_1 + 2 X_3$ &  10 & $3 X_1 + 3 X_3$ &  12 & $5 X_1 + 4 X_3 + X_5$\\ 
\hline
\end{tabular}
\end{table}

It is interesting to briefly focus on specific multipartite entangled 
four qubit states. One prominent class of states is formed by the so called 
{\em graph states}~\cite{briegel01,hein04}.
They are created from a fully polarized state in e.g. $x$-direction by
successive action of the two-qubit entangling operator 
$U_{ij}:=\frac{1}{2}(\id+\sigma_{3;i}+\sigma_{3;j}-\sigma_{3;i}\sigma_{3;j})$ 
which is also known as the control-$\sigma_3$ gate.
A complete characterization of graph states for up to seven qubits 
can be found in Ref.~\cite{hein04}.
In the case of four qubits, only two graph state classes exist. 
Representatives are the GHZ state and the 
4-qubit cluster state~\cite{briegel01}. A genuinely entangled
four qubit state that falls out of this classification, 
namely $\ket{X}:=\frac{1}{\sqrt{6}}
    (\sqrt{2}\ket{1111}+\ket{1000}+\ket{0100}+\ket{0010}+\ket{0001})$, 
has been presented in Ref.~\cite{OS04,OS05}
together with an evaluation of the filters ${\cal F}^{(4)}_1$, ${\cal F}^{(4)}_2$,
and ${\cal F}^{(4)}_3$ on these three states. 
It is worth noticing that the three filters admit for
$7=2^3-1$ classes of genuine four-party entanglement (in the nonzero SLOCC class).
Representative states for these seven classes of entanglement
can be obtained as coherent superpositions of GHZ, cluster, and X state from the
table of filter values in Ref.~\cite{OS05}. In this sense these three 
maximally entangled states form a basis for the whole nonzero SLOCC class.
Consequently, a classification
of genuine multipartite entanglement in terms of graph states alone 
is not complete. 
The account of the complementary set of non-graph states as a resource for 
quantum information processing is largely unexplored. 

\section{$SL$ and $SL^*$-invariants for five qubits}
\label{fivebits}

The $SL$ Hilbert series for five qubits has been determined in Ref.~\cite{Luque05}
as
\beqa\label{HilbertSeries}
h(t)&=&
\frac{1+16 t^8 + 9 t^{10}+82 t^{12} + \dots + 82 t^{92}+ 9 t^{94}+ 16 t^{96}+ t^{104}}{
(1-t^4)^5(1-t^6)(1-t^8)^5(1-t^{10})(1-t^{12})^5}\\
&=& 1 + 5 t^4 + t^6 + 36 t^8 + 15 t^{10} + 
 228 t^{12} +231 t^{14} + 1313 t^{16} + 1939 t^{18}\nonumber\\
&& + \dots
\label{ExpandedHilbertSeries}
\eeqa
We have verified the values of the coefficients $c_{2d}$ of $t^{2d}$ in 
Eq.~\eqref{ExpandedHilbertSeries} by using the formula
$c_{2d} = (1/(2d)!) \sum_{\pi\in S_{2d}} \chi(\pi)^5$
where $\chi(\pi)$ is the character of the irreducible representation of 
$S_{2d}$ corresponding to the partition $[d,d]$ of the integer $2d$.
This is a special case of the formula from Eq.~\eqref{dim1} 
where we replace $d$ with $2d$ and insert the local Hilbert space dimension 
$n=2$ and the number of qubits $k=5$.
Both numerator and denominator in Eq.~\eqref{HilbertSeries} 
are even palindromic polynomials of degrees $104$ and $136$ respectively.
The expanded Hilbert series tells us that 
there are $5$ invariants of degree $4$,
a single invariant of degree $6$, $36$ invariants of degree $8$,
$15$ invariants of degree $10$, $228$ invariants of degree $12$, etc.
In Ref.~\cite{Luque05} the invariants up to degree $6$ have been determined
together with $5$ invariants of degree $8$. 

The first terms of the $SL^*$ Hilbert series are (for details see Section~\ref{SLstarseries})
$$ h_{SL^*}(t) = 1+t^4+4t^8+12 t^{12}+2 t^{14}+39 t^{16}
+21 t^{18}+130 t^{20}+115 t^{22}+\dots\; . $$

In this section we will give a complete characterization
of invariants up to degree $12$ and establish a connection with 
the invariants from Ref.~\cite{OS05}. 
Since the Hilbert series shows that no invariant of degree $2$ 
exists, we start our analysis with degree $4$.

\subsection{Degree $4$}
A straightforward calculation shows that the $5$ linearly independent
invariants $D_v$ of degree $4$ ($v=x,y,z,t,u$) from Ref.~\cite{Luque05}
can be written as
\beqa\label{D1}
D_1:=D_x&=&(\!(\sigma_\mu\sigma_2\sigma_2\sigma_2\sigma_2
    \bullet \sigma^\mu\sigma_2\sigma_2\sigma_2\sigma_2)\!) \\
D_2:=D_y&=&(\!(\sigma_2\sigma_\mu\sigma_2\sigma_2\sigma_2
    \bullet \sigma_2\sigma^\mu\sigma_2\sigma_2\sigma_2)\!)\label{D2} \\
D_3:=D_z&=&(\!(\sigma_2\sigma_2\sigma_\mu\sigma_2\sigma_2
    \bullet \sigma_2\sigma_2\sigma^\mu\sigma_2\sigma_2)\!)\label{D3} \\
D_4:=D_t&=&(\!(\sigma_2\sigma_2\sigma_2\sigma_\mu\sigma_2
    \bullet \sigma_2\sigma_2\sigma_2\sigma^\mu\sigma_2)\!) \label{D4}\\
D_5:=D_u&=&(\!(\sigma_2\sigma_2\sigma_2\sigma_2\sigma_\mu
    \bullet \sigma_2\sigma_2\sigma_2\sigma_2\sigma^\mu)\!) \label{D5}\; .
\eeqa
Interestingly, these invariants and their generalizations to higher odd
number of qubits already appeared in Ref.~\cite{Wong00}.
They form an $S_5$-orbit, 
which is nicely seen from their explicit forms \eqref{D1}--\eqref{D5}. 
The unique $SL^*$-invariant of degree $4$ is then 
$$
P:=\sum_{i=1}^5 D_i\; .
$$
It does not vanish on all product states. Therefore,
${\cal I}^{SL^*}_{0;4}=0$ for five qubits.
It is not obvious whether this observation can be extended 
to larger number of qubits.

The investigation of the full $S_5$-orbits of a given set of invariants
will be a major tool for the construction of the complete
space of invariants and the determination of the $SL^*$-invariants. 
In the present case, we only needed e.g. $D_1$
in order to create all degree $4$ invariants from its orbit.
The decomposition into irreducible $S_5$-modules is $V_4=X_1+X_2$.

It is an interesting consequence of the completeness of \eqref{D1}--\eqref{D5}
as generators of invariants of degree $4$ that
additional contractions lead to identities. Two examples are
\beqa
(\!(\sigma_\mu \sigma_\nu \sigma_\lambda \sigma_2 \sigma_2
\bullet \sigma^\mu \sigma^\nu \sigma^\lambda \sigma_2 \sigma_2)\!)
&=&3(D_4+D_5)-P,\label{mmm22Id} \\
(\!(\sigma_\mu \sigma_\nu \sigma_\lambda \sigma_\tau \sigma_\rho
\bullet \sigma^\mu \sigma^\nu \sigma^\lambda \sigma^\tau \sigma^\rho)\!)
&=& -3 P\; , \label{PsymmId}
\eeqa
but also note the above mentioned identities for four qubits.
Up to the prefactor, equation~\eqref{PsymmId} readily follows from the 
obvious permutation symmetry.

\subsection{Degree $6$ and singly even $q$}
The unique invariant $F$ of degree $6$ has been created in Ref.~\cite{Luque05} 
by invoking the $\Omega$-process.
It is an odd function under qubit permutations, corresponding to the 
irreducible $S_5$-module $V_6=X_7$. It cannot
be created from the combs. However, see section~\ref{altomega}. 
Remembering degree $2$, where we
know from the Hilbert series that ${\rm Inv}^{SL}_2=0$ for an odd numer of qubits,
it seems that singly even $q$ is peculiar. 
This is particularly true as far as the comb-based method is concerned. 
The only invariant of degree $2$ that can be created from 
local invariant operators is the $q$-tangle~\cite{Wong00}
$(\!(\sigma_y^{\otimes q})\!)$, which however is identically zero when $q$ is odd. 
This phenomenon draws wider circles, as expressed in 
\begin{theorem}
For an odd number of qubits, all nonzero $SL$-invariants that can be 
constructed from combs have doubly even degree. 
\end{theorem}

{\em Proof: }An expectation value
$(\!(\sigma_{a_1}\sigma_{a_2}\dots)\!)$ vanishes if it contains 
an odd number of $\sigma_2$'s~\cite{OS05}. Since the contraction with the 
pseudo-metric $G_{\mu\nu}$ does not contain $\sigma_2$ this implies
that for an odd number of qubits there must be an odd number of contractions
in each copy, leading to an odd
number of contractions where all contractions in each of the copies 
are counted separately. 
Since the copies are always contracted in pairs, this number of contractions must
be even. Therefore, the number of copies
for an odd number of qubits has to be even,
leading to a degree divisible by $4$, hence doubly even.
This completes the proof.

It seems that this fact is intimately related to the observed permutation
antisymmetry of the invariants of singly even degree (we anticipate here that
the one-dimensional $S_5$-modules of the generic invariants of degree $6$ and $10$
are spanned by odd functions under qubit permutations).
The symmetry of the combs under permutations of the copies might hinder
assymmetry under qubit exchange, even though
there is a profound difference between copy and qubit exchange. 
Indeed, as we will see later, nonzero antisymmetrizations of comb-based invariants do exist.
Nevertheless, it is natural to ask for local invariants
that are antisymmetric under the permutation of copies;
it turns out that no such construction exists that connects
two or three copies, i.e. there are no antisymmetric combs of 
order two or three. 
Also notice that no independent symmetric combs exist
up to degree four \footnote{A detailed presentation of this result 
will appear elsewhere.}.

\subsection{Degree $8$}
We next proceed with a complete discussion of degree $8$ invariants.  
Looking at the Hilbert series, the dimension of this space is $36$. 
The $15$ products $D_iD_j$, $1\le i\le j\le 5$, form a basis of the subspace $U_8$. 
This implies the existence of $21$ independent invariants that are
generically of degree $8$. Five of these have been constructed in 
Ref.~\cite{Luque05} by using the $\Omega$-process.
We will at first give the decomposition of $U_8$ and $V_8$ into irreducible
$S_5$-modules and then establish the connection with   
these five invariants $H_v$.
We find that
\beqa\label{S5-8ModulesU}
U_8 &=& 2X_1 + 2X_2 + X_3, \\
V_8 &=& 4 X_1 + 3  X_2 + 3 X_3 + X_5. \label{S5-8ModulesV}
\eeqa
The dimension of ${\rm Inv}^{SL^*}_8$ for $5$ qubits is consequently $4$,
which agrees with the $SL^*$ Hilbert series.

The filter 
\beqa
{\cal F}^{(5)}_1 &=&(\!(\sigma_{\mu_1}\sigma_{\mu_2}\sigma_{\mu_3}\sigma_2\sigma_2
	      \bullet \sigma^{\mu_1}\sigma^{\mu_2}\sigma_2\sigma_{\mu_4}\sigma_2\bullet\nonumber\\
&& \qquad \sigma_{\mu_5}\sigma_2\sigma^{\mu_3}\sigma^{\mu_4}\sigma_2
 \bullet \sigma^{\mu_5}\sigma_2\sigma_2\sigma_2\sigma_2)\!)\label{F1}
\eeqa
has been introduced in Ref.~\cite{OS05}. We add two new $SL$-invariants 
\beqa
{\cal F}^{(5)}_5 &=& 
(\!(\sigma_\mu \sigma_2  \sigma_2  \sigma_\nu  \sigma_\lambda\bullet    
\sigma^\mu \sigma_\rho \sigma_2 \sigma_2  \sigma^\lambda\bullet    \nonumber\\
&&\qquad \sigma_2 \sigma^\rho \sigma_\tau\sigma_2\sigma_\kappa\bullet    
 \sigma_2   \sigma_2   \sigma^\tau \sigma^\nu \sigma^\kappa)\!), \label{F2}\\
{\cal F}^{(5)}_6 &=& 3
(\!(\sigma_\mu \sigma_\nu \sigma_\lambda \sigma_2 \sigma_2 \bullet        
\sigma_\tau \sigma^\nu \sigma^\lambda \sigma_2 \sigma_2\bullet    \nonumber\\ 
&&\qquad \sigma^\tau\sigma_2\sigma_2\sigma_\rho \sigma_\kappa  \bullet      
\sigma^\mu  \sigma_2 \sigma_2 \sigma^\rho \sigma^\kappa)\!) 
\nonumber   \\
         &&+ \label{T}\\
&& (\!(\sigma_\mu \sigma_\nu \sigma_\lambda \sigma_2 \sigma_2\bullet    
\sigma^\mu \sigma^\nu \sigma^\lambda \sigma_2 \sigma_2\bullet    \nonumber\\
&&\qquad \sigma_\tau \sigma_2 \sigma_2 \sigma_\rho \sigma_\kappa\bullet    
\sigma^\tau \sigma_2 \sigma_2\sigma^\rho \sigma^\kappa)\!). \nonumber
\nonumber
\eeqa
Notice that the second summand in \eqref{T} is an element of $U_8$.
More precisely, $(\!(\sigma_\mu \sigma_\nu \sigma_\lambda \sigma_2 \sigma_2
\bullet \sigma^\mu \sigma^\nu \sigma^\lambda \sigma_2 \sigma_2)\!)=3(D_4+D_5)-P$.

We claim that these two invariants are filters.
It is straightforward to check this claim for the invariant \eqref{F2}.
To see that also \eqref{T} is a filter it suffices to show that it
vanishes on product states. The only partitions 
that lead to a nonzero value for both terms in the above sum are
those factoring out either qubits $(2,3)$ or qubits $(4,5)$.
The nonzero value is a multiple of powers of
concurrence and three-tangle (e.g. $C_{(2,3)}^4 \tau_{3;(1,4,5)}^2$), 
and the prefactor is found to be independent of which of the two distinct
partitions we take. 
It is then straightforward to check
that the above combination vanishes also for these factorizations,
proving the filter property.

We will show next that, together with $P^2$, the permutation averages of
the above three filters span the space ${\rm Inv}^{SL^*}_8$.
The $S_5$-submodule generated by the filter ${\cal F}^{(5)}_1$ has
dimension 24 and meets $U_8$ in an $X_2$, a $4$-dimensional subspace.
Thus by selecting 20 suitable qubit permutations of this filter, we obtain
altogether $15+20=35$ linearly independent invariants in $V_8$.
To obtain a basis of $V_8$, we have to add also the filter
${\cal F}^{(5)}_6$.

It is interesting that the filter \eqref{F1} resp. \eqref{F2} creates a 
$24$-dimensional space of invariants $V_{8;1}$ resp. $V_{8;2}$. These two
spaces have a $23$-dimensional overlap $K$. Thus
$V_{8;i}=T_i+K$; $i=1,2$, where $T_i$ are one-dimensional subspaces
of ${\rm Inv}^{SL^*}$. 
Furthermore, also the space created from \eqref{T}, which we will call 
$V_{8;3}$ can be expressed as $V_{8;3}=T_3+\kappa$, where $\kappa\subset K$ 
and the subspace $T_3\subseteq{\rm Inv}^{SL^*}$ is one-dimensional.
Since these spaces have been created from filters, the $T_i$ ($i=1,2,3$)
are already the elements in  ${\cal I}^{SL^*}_0$ we have been looking for.
These particular invariants are given by
$$
T_{1;0}:=\expect{{\cal F}^{(5)}_1}_s, \
T_{2;0}:=\expect{{\cal F}^{(5)}_5}_s, \
T_{3;0}:=\expect{{\cal F}^{(5)}_6}_s.
$$
A detailed analysis of the characters of the resulting irreducible
$S_5$-modules leads to the decomposition \eqref{S5-8ModulesV}.

We now give the expression of the three invariants in ${\cal I}^{SL^*}_0$ 
in terms of the invariants obtained in Ref.~\cite{Luque05}.
To this end, we define a second $SL^*$-invariant in $U_8$, namely
$$
Q:=\sum_{i=1}^5 D_i^2 
$$
and use the sum of all the $H_v$
$$
H_0:=\sum_{i=1}^5 H_i\ .
$$
We find that
$$
T_{2;0}=P^2 - 3 Q \in U_8, \quad T_{3;0}=H_0 + P^2 -6 Q.
$$
Summarizing, ${\rm Inv}^{SL^*}_8$ is spanned by $P^2,T_{1;0},T_{2;0},T_{3;0}$, or 
equivalently by $P^2,Q,T_{1;0},T_{3;0}$. The subspace 
${\cal I}^{SL^*}_{0;8}$ is spanned by $T_{1;0},T_{2;0},T_{3;0}$.

\subsection{Degree $10$}

From the Hilbert series we extract that there are $15$ independent 
invariants of degree $10$, where $5$ independent elements of
$U_{10}$ are obtained by multiplying the $5$ invariants $D_i$, 
$i=1,\dots,5$ with the invariant $F$ of degree $6$. 
Hence, $U_{10}=X_6+X_7$ as an 
$S_5$-module \footnote{We follow the notation of Ref.~\cite{Luque05}.}.

The missing ten invariants are in the $S_5$-orbit of ${\cal G}^{(5)}_{10}$ 
which we construct by invoking the following $\Omega$-process
(for the notation used here see Section \ref{altomega})
\beqa
B_{00222}&:=&(f,f)^{11000}\nonumber \\
B_{20022}&:=&(f,f)^{01100}\nonumber  \\
B_{20202}&:=&(f,f)^{01010}\nonumber  \\
C_{20222}&:=&(B_{20022},B_{00222})^{00011}\nonumber  \\
D_{11131}&:=&(C_{20222},f)^{10101}\nonumber  \\
E_{20222}&:=&(D_{11131},f)^{01010}\nonumber  \\
F_{11311}&:=&(E_{20222},f)^{10011}\nonumber  \\
H_{11111}&:=&(F_{11311},B_{20202})^{10201}\nonumber  \\
{\cal G}^{(5)}_{10}&:=&(H_{11111},f)^{11111}\label{omega10}
\eeqa
${\cal G}^{(5)}_{10}$ spans a $14$-dimensional space which has a $4$-dimensional
intersection with $U_{10}$.
In terms of irreducible $S_5$-modules the space of degree $10$ invariants
decomposes as
$$
V_{10}=X_5 + 2 X_6 + 2 X_7
$$
with dimension counting $15=5+2*4+2*1$. This shows that there are no
$SL^*$-invariants; however, there are two odd symmetric 
invariants: $P\cdot F \in U_{10}$ and the anti-symmetrization of
${\cal G}^{(5)}_{10}$. Both are in the ideal ${\cal I}^{SL^*_{-}}_0$.

Summarizing, 
${\rm Inv}^{SL^*_{-}}_{10}={\cal I}^{SL^*_{-}}_{0;10}={\rm span} 
\{P\cdot F,\left\langle{\cal G}^{(5)}_{10}\right\rangle_a\}$.

\subsection{Degree $12$}

From the Hilbert series we see that the space $V_{12}$ of degree $12$ 
invariants has dimension $228$, where a $141$-dimensional subspace $U_{12}$
emerges from lower degrees. The latter space decomposes as
$$
U_{12} = 7 X_1 + 10 X_2 + 8 X_3 + 5 X_4 + 4 X_5 + X_6.
$$
Hence there are $87$ invariants that are generically of degree $12$.
For the complete reconstruction and decomposition of this space into
irreducible $S_5$-modules we use the filters
rather than employing the $\Omega$-process, since this reduces significantly the
computational complexity. The origin of this reduction in computational complexity 
can be understood from the analysis in Section~\ref{altomega}.

We claim that the $S_5$-orbits of the five invariants
\beqa
{\cal F}^{(5)}_{12;1} 
&=&(\!(\sigma_{\mu_1}\sigma_{\mu_2}\sigma_{\mu_3}\sigma_2\sigma_2\bullet    
		  \sigma^{\mu_1}\sigma_2\sigma_2\sigma_{\mu_4}\sigma_{\mu_5}\bullet    
\label{5-12-1}\\
&& \qquad \sigma_2\sigma^{\mu_2}\sigma_2\sigma_2\sigma_2
 \bullet \sigma_2\sigma_2\sigma^{\mu_3}\sigma_2\sigma_2 \bullet    \nonumber \\
&& \qquad \sigma_2\sigma_2\sigma_2\sigma^{\mu_4}\sigma_2
 \bullet \sigma_2\sigma_2\sigma_2\sigma_2\sigma^{\mu_5})\!)\nonumber \\
{\cal F}^{(5)}_{12;2} &=&(\!(\sigma_{\mu_1}\sigma_{\mu_2}\sigma_{\mu_3}\sigma_2\sigma_2
	    \bullet\sigma^{\mu_1}\sigma^{\mu_2}\sigma_{\mu_4}\sigma_2\sigma_2\bullet
\label{5-12-2} \\ 
&& \qquad  \sigma_{\mu_5}\sigma_2\sigma^{\mu_3}\sigma_2\sigma_{\mu_6}\bullet    
	     \sigma^{\mu_5}\sigma_2\sigma^{\mu_4}\sigma_2\sigma_{\mu_7}\bullet\nonumber\\
&& \qquad \sigma_{\mu_8}\sigma_2\sigma_2\sigma_{\mu_9}\sigma^{\mu_6}\bullet
		  \sigma^{\mu_8}\sigma_2\sigma_2\sigma^{\mu_9}\sigma^{\mu_7})\!)\nonumber\\
{\cal F}^{(5)}_{12;4} &=& 
(\!(\sigma_{\mu_1} \sigma_{\mu_2}\sigma_{\mu_3}\sigma_2\sigma_2 \bullet
\sigma^{\mu_1}  \sigma_2\sigma_{\mu_4}\sigma_{\mu_5}\sigma_2 \bullet
\label{5-12-4}\\
&& \qquad  \sigma_2 \sigma^{\mu_2}\sigma^{\mu_3}\sigma_2\sigma_{\mu_6}\bullet 
\sigma_{\mu_7}  \sigma_2\sigma^{\mu_4}\sigma_2 \sigma^{\mu_6}\bullet \nonumber\\
&& \qquad \sigma^{\mu_7} \sigma_2\sigma_2 \sigma^{\mu_5}\sigma_{\mu_8}\bullet 
\sigma_2 \sigma_2 \sigma_2 \sigma_2 \sigma^{\mu_8})\!)\nonumber\\
{\cal G}^{(5)}_{12;2} &=&
(\!(\sigma_{\mu_1}\sigma_{\mu_2}\sigma_{\mu_3}\sigma_{2} \sigma_{2}\bullet 
\sigma^{\mu_1}\sigma^{\mu_2} \sigma_{\mu_4} \sigma_2\sigma_2 \bullet 
\label{5-12-G2}\\
&& \qquad \sigma_{\mu_5}\sigma_{\mu_6}\sigma^{\mu_3}\sigma_2 \sigma_2 \bullet 
  \sigma^{\mu_5} \sigma_{\mu_7} \sigma_{2}\sigma_{\mu_8}\sigma_2\bullet  \nonumber\\
&& \qquad \sigma_{\mu_9}\sigma^{\mu_6} \sigma_2 \sigma^{\mu_8}\sigma_2 \bullet 
\sigma^{\mu_9} \sigma^{\mu_7}\sigma^{\mu_4}\sigma_2 \sigma_{2})\!)\nonumber \\
{\cal G}^{(5)}_{12;6} &=&
(\!(\sigma_{\mu_1}\sigma_{\mu_2}\sigma_{\mu_3}\sigma_{\mu_4} \sigma_{\mu_5}\bullet 
\sigma^{\mu_1}\sigma_2 \sigma_2 \sigma_2\sigma_2 \bullet 
\label{5-12-G6}\\
&& \qquad\sigma_2\sigma^{\mu_2}\sigma_2\sigma_{\mu_6}\sigma_{\mu_7} \bullet  
\sigma_{\mu_8}\sigma_{\mu_9}\sigma^{\mu_3}\sigma^{\mu_6}\sigma^{\mu_7}\bullet \nonumber\\
&& \qquad\sigma^{\mu_8}\sigma^{\mu_9} \sigma_2 \sigma^{\mu_4}\sigma_2 \bullet 
\sigma_{2} \sigma_{2}\sigma_2\sigma_2 \sigma^{\mu_5})\!)\nonumber
\eeqa
generate the full set of $87$ new invariants.
The symbol ${\cal F}$ indicates that the invariant has the filter property.
The filters \eqref{5-12-1} and \eqref{5-12-2} are taken from ~\cite{OS05},
whereas the invariants \eqref{5-12-4}, \eqref{5-12-G2} and \eqref{5-12-G6} are new.

We start our construction of a basis of $V_{12}$
by choosing a basis of the subspace $U_{12}$ (141 elements).
Next we make use of the filter ${\cal F}^{(5)}_{12;4}$. The $S_5$-module
that it generates has dimension 112 and intersects $U_{12}$ in a
$44$-dimensional submodule. Thus we can construct the next 68 basis
elements by applying suitable qubit permutations to this filter.
The next 15 elements of the basis are obtained similarly from
${\cal G}^{(5)}_{12;2}$, and 2 more from ${\cal F}^{(5)}_{12;2}$.
This gives in total $141+68+15+2=226$ basis elements. A full basis of $V_{12}$
is obtained by adjoining the invariants ${\cal F}^{(5)}_{12;1}$ and 
${\cal G}^{(5)}_{12;6}$. This proves the claim made above.

It is staightforward to construct filters from ${\cal G}^{(5)}_{12;2}$ and
${\cal G}^{(5)}_{12;6}$ by subtracting suitable elements of $U_{12}$. 
In both cases there is a single partition
for which the invariant does not vanish on corresponding product states: 
for the partition $(1,2)(3,4,5)$ we have 
${\cal G}^{(5)}_{12;2}=9 C_{1,2}^6\cdot\tau_{3;3,4,5}^3$ 
whereas for $(2,4)(1,3,5)$ we obtain that 
${\cal G}^{(5)}_{12;6}=3 C_{2,3}^6\cdot\tau_{3;1,4,5}^3$.
Filters are constructed by subtracting the $U_{12}$-elements
\nbeqa
\Delta{\cal G}^{(5)}_{12;2} &=&
(\!(\sigma_{\mu_1}\sigma_{\mu_2}\sigma_{\mu_3}\sigma_{2} \sigma_{2}\bullet 
\sigma^{\mu_1}\sigma^{\mu_2} \sigma^{\mu_3} \sigma_2\sigma_2 )\!)\cdot \\
&& \qquad (\!(\sigma_{\mu_5}\sigma_{\mu_6}\sigma_{\mu_4}\sigma_2 \sigma_2 \bullet 
  \sigma^{\mu_5} \sigma_{\mu_7} \sigma_{2}\sigma_{\mu_8}\sigma_2\bullet  \\
&& \qquad \sigma_{\mu_9}\sigma^{\mu_6} \sigma_2 \sigma^{\mu_8}\sigma_2 \bullet 
\sigma^{\mu_9} \sigma^{\mu_7}\sigma^{\mu_4}\sigma_2 \sigma_{2})\!),  \\
\Delta{\cal G}^{(5)}_{12;6} &=&-\frac{1}{9}
(\!(\sigma_{\mu_1}\sigma_{\mu_2}\sigma_{\mu_3}\sigma_{\mu_4} \sigma_{\mu_5}\bullet 
\sigma^{\mu_1}\sigma^{\mu_2}\sigma^{\mu_3}\sigma^{\mu_4}\sigma^{\mu_5})\!)\cdot \\
&& \qquad(\!(\sigma_2\sigma_{\mu_6}\sigma_2\sigma_{\mu_7}\sigma_{\mu_8} \bullet  
\sigma_{\mu_9}\sigma^{\mu_6} \sigma_2 \sigma^{\mu_7}\sigma_2 \bullet \\
&&\qquad \sigma^{\mu_9}\sigma_2 \sigma_2 \sigma_2\sigma_2 \bullet 
\sigma_{2} \sigma_{2}\sigma_2\sigma_2 \sigma^{\mu_8})\!)
\neeqa
from ${\cal G}^{(5)}_{12;2}$ and ${\cal G}^{(5)}_{12;6}$, respectively.

It is interesting to mention here that an $SL^*$-filter
can readily be constructed from combs as follows
\nbeqa
{\cal F}^{(5)}_0 &=&(\!(\sigma_{\mu_1}\sigma_{\mu_2}\sigma_{\mu_3}\sigma_{\mu_4}\sigma_{\mu_5}
		  \bullet \sigma^{\mu_1}\sigma_2\sigma_2\sigma_2\sigma_2\bullet \\
&& \qquad \sigma_2\sigma^{\mu_2}\sigma_2\sigma_2\sigma_2
 \bullet \sigma_2\sigma_2\sigma^{\mu_3}\sigma_2\sigma_2\bullet \\
&& \qquad \sigma_2\sigma_2\sigma_2\sigma^{\mu_4}\sigma_2
 \bullet \sigma_2\sigma_2\sigma_2\sigma_2\sigma^{\mu_5})\!) 
\neeqa
It turns out that ${\cal F}^{(5)}_0$ is equivalent modulo $U_{12}$
to the symmetrization of ${\cal F}^{(5)}_{12;1}$.

We find that the $7$-dimensional space $U_{12}\cap{\rm Inv}^{SL^*}_{12}$ 
is spanned by $P^3$, $PT_{j;0}$ (j=1,2,3), $F^2$, $\sum_i D_i^3$, 
and $\expect{D_1 {\cal F}^{(5)}_1}_s$. 
Besides $PT_{j;0}$ (j=1,2,3), $\expect{D_1 {\cal F}^{(5)}_1}_s$ 
and ${\cal F}^{(5)}_0$, also
$P^3- 9 \sum_i D_i^3$ is in ${\cal I}^{SL^*}_0$.

The complementary $5$-dimensional space in ${\rm Inv}_{12}^{SL^{*}}$ is
spanned by
$$
\expect{{\cal F}^{(5)}_{12;1}}_s\; ,\quad
\expect{{\cal F}^{(5)}_{12;2}}_s \; ,\quad
\expect{{\cal F}^{(5)}_{12;4}}_s\; ,\quad
\expect{{\cal G}^{(5)}_{12;2}}_s\; ,\quad
\expect{{\cal G}^{(5)}_{12;6}}_s\; .
$$
The two antisymmetrized filters
$$
\expect{{\cal F}^{(5)}_{12;2}}_a\; , \quad
\expect{{\cal F}^{(5)}_{12;4}}_a\; 
$$
span the space of $SL^*_-$-invariants of degree $12$,
whereas
$\expect{3 {\cal G}^{(5)}_{12;6}- {\cal G}^{(5)}_{12;2}}_s$
is in ${\cal I}^{SL^*}_0$.
It is worthwhile noticing that the comb-based invariants 
also create the $SL^*_-$-invariants in $U_{12}$; those of degree $6$ and $10$,
which are not accessible by the comb approach, are not needed. 

Summarizing we have a $10$-dimensional space ${\cal I}^{SL^*}_{0;12}$
inside a $12$-dimensional space ${\rm Inv}^{SL^*}_{12}$.
In addition there are two $SL^*_-$-invariants, both belong to
the filter ideal. Thus,
\nbeqa
{\cal I}^{SL^*}_{0;12}&=&{\rm span}\left\{P\cdot T_{j,0}|_{j=1,\dots ,3},F^2,
\left\langle{\cal F}^{(5)}_1\right\rangle_s,{\cal F}^{(5)}_0,P^3
-9\sum_i D_i^3,\left\langle 3{\cal G}^{(5)}_{12,6}- 
{\cal G}^{(5)}_{12,2}\right\rangle_s\right\}\\
{\cal I}^{SL^*_{-}}_{0;12}&=&{\rm span}\left\{
\left\langle{\cal F}^{(5)}_{12;2}\right\rangle_a,
\left\langle{\cal F}^{(5)}_{12;4}\right\rangle_a\right\}\ . 
\neeqa
The complete decomposition into irreducible $S_5$-modules is given in
table \ref{inv5}.

\begin{table}[h]
\caption{The space of polynomial invariants of degree $2$ up
to $12$ into irreducible $S_5$-modules.}\label{inv5}
\begin{tabular}{|c|l||c|l|}
\hline
degree & & degree & \\
\hline
\hline
2 &  $ 0 $ &
4 & $X_1 + X_2$ \\ \hline  
6 & $X_7$ &
8 & $4X_1 + 3 X_2 + 3 X_3 + X_5$ \\ \hline  
10 & $X_5 + 2 X_6 + 2 X_7$ &  12 & 
$\scriptstyle 12 X_1 + 15 X_2 + 14 X_3 + 6 X_4 + 8 X_5 + 2 X_6 + 2 X_7$\\
\hline
\end{tabular}
\end{table}

\subsection{Beyond degree $12$}

We add here a couple of remarks about the next two degrees, $14$ and $16$.
To this end, let $Y$ denote a minimal set of (homogeneous)
generators of the algebra ${\rm Inv}^{SL}$.
We know that $Y$ is a finite set, but its
cardinality is not known. It is a disjoint union of the
subsets $Y_d:=Y \cap {\rm Inv}_d^{SL}$.
From the Hilbert series we know that
$|Y_d|=0$ for odd $d$ and for $d=2$.
Our computations show that for $d=4,6,8,10,12,14,16$
we have $|Y_d|=5,1,21,10,87,145,247$.

Let us assume that the conjecture made in Ref.~\cite{Luque05}
regarding the Cohen--Macaulay ring structure of
${\rm Inv}^{SL}$ is correct, i.e., that the primary
invariants consist of five polynomials of degree 4,
one of degree 6, five of degree 8, one of degree 10,
and five of degree 12. Then ${\rm Inv}^{SL}$ would
be a free module of rank 3 014 400 over the
algebra generated by the primary invariants
(a polynomial algebra in 17 variables). Moreover,
the coefficients of the numerator of the Hilbert
series give, for each degree, the number of basis
elements of this free module. The first six nonzero coefficients
are $1,16,9,82,145,383$ and
the degrees of the corresponding basis elements
are $0,8,10,12,14,16$, respectively
(see Table 1 in Ref.~\cite{Luque05}).
For instance, for $d=16$ we have 383 basis elements. 
We may assume that this basis contains $Y_{16}$.
Consequently there must be $383-247=136$ basis
elements of degree 16 that come from the products
of basis elements of degree 8. As there are sixteen
basis elements of degree 8, the number of different
products of two of them (including the squares)
is indeed 136. This may be interpreted as additional
evidence for the validity of the above mentioned conjecture.

\begin{figure}[h]
\includegraphics[width=.8\textwidth]{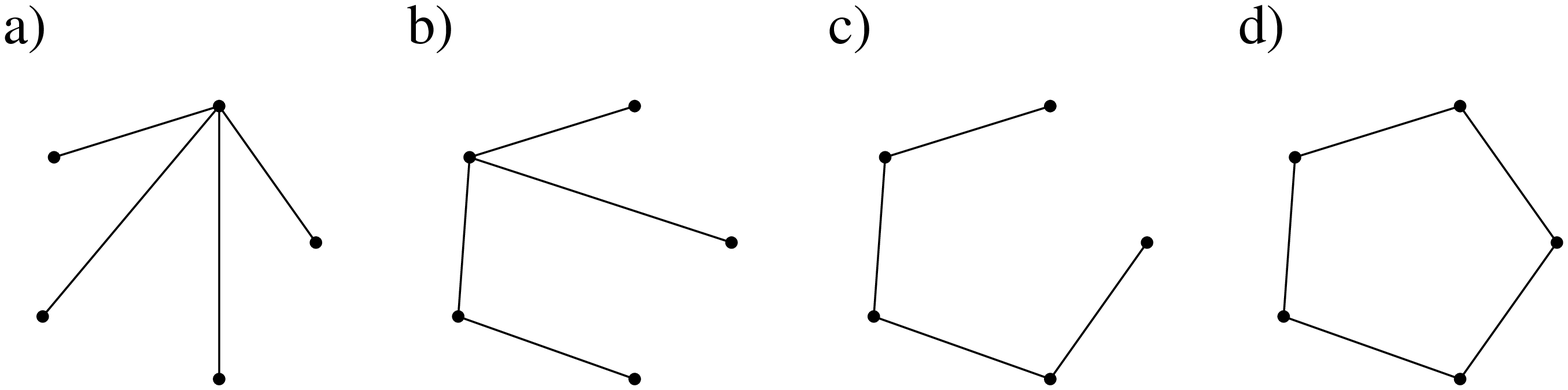}
\caption{Full set of inequivalent graph states on $5$ qubits as graphically
depicted in Ref.~\cite{hein04}. 
a) is equivalent to
the GHZ state $\ket{0}+\ket{31}$, where the binary decomposition of the numbers 
gives the qubit state, e.g. $\ket{31}\equiv\ket{11111}$. b) is equivalent to
$\ket{1}+\ket{2}+\ket{28}+\ket{31}$ and emerges from the $4$-qubit cluster state
after copying and stacking one qubit (telescoping). c) is equivalent to 
$\ket{1}+\ket{6}+\ket{24}+\ket{31}$ and emerges from the $3$-qubit GHZ state 
$\ket{1}+\ket{2}+\ket{4}+\ket{7}$ after two-qubit telescoping. 
Finally, d) is the superposition of two orthogonal and telescoped $4$-qubit cluster states.
They are distinguished already by the invariants $D_i$ of degree $4$.
However, all $D_i$ vanish for state d). The $5$-tuple of 
(relative) $SL^*$ generators up to degree 
$8$ -- $(P,F,T_{1;0},T_{2;0},T_{3;0})$ -- evaluated on these four states gives
a) $(5,0,-1,1,4)$, $\mbox{b)  $(2,0,\frac{1}{10},-\frac{1}{5},-\frac{4}{5})$}$, 
c)  $(-1,0,\frac{1}{15},-\frac{1}{5},\frac{8}{5})$, and d) $(0,0,0,0,6)$.
\label{5-bit-graphs}}
\end{figure}

It is interesting to examine again the class of graph states.
For five qubits there exist four inequivalent graph states: numbers $5$ through $8$ in 
Ref.~\cite{hein04} (see figure \ref{5-bit-graphs}). 
They are distinguished already by the degree $4$ invariants 
$D_i,\; i=1,\dots\, 5$. Among $SL^*$-invariants up to degree $8$,
the state d) of figure~\ref{5-bit-graphs} is only detected
by $T_{0;3}$ (and hence also by $H_0$).
It is worth mentioning that the $SL^*_-$-invariant $F$ does not 
detect any of these states. 
Besides the two maximally entangled states from Ref.~\cite{OS05}
$\ket{1}+\ket{2}+\ket{4}+\ket{24}+\sqrt{2}\ket{31}$ and
$W_4 + \sqrt{3}\ket{31}$, also superpositions of these two states 
and possibly including the four graph states fall out of the 
graph state classification. 
This basis of $6$ states would thereby admit for already $2^6-1=63$ distinct
SLOCC classes of $5$-qubit states. Only four of these are covered by 
graph states.

\section{Character computations and the Hilbert series for $SL^*_5$}
\label{SLstarseries}

It is interesting to mention that the coefficients of the
Hilbert series for the symmetry group $SL$ and $SL^*$ 
can be obtained directly using the results of Ref.~\cite{Brylinski02b}. 
Here, we recall some results from that work and
use them to compute the dimension of the space of $SL^*$-invariants
of degree $2d$. We also do the same for the $SL^*_-$ invariants.

As in the cited reference, we shall be more general. First, instead
of qubits we may work with qudits, i.e., we consider the
vector representation of $SU(n)$ or $SL(n,\CC)$ on $V=\CC^n$.
By taking $k$ copies of $SL(n,\CC)$ and $k$ copies of $V$ and tensoring,
we obtain the standard representation of $G=SL(n,\CC)^{\otimes k}$
on $V^{\otimes k}$. Let us denote by $\pR_{n,k}$ the algebra of 
holomorphic polynomial functions on $V^{\otimes k}$, and by
$\pR_{n,k,d}$ its subspace consisting of the homogeneous polynomials
of degree $d$. Next, denote by $\pR_{n,k,d}^G$ the subspace of 
$\pR_{n,k,d}$ consisting of $G$-invariant polynomials. 
If $d$ is not divisible by $n$, then $\pR_{n,k,d}^G=0$
by \cite[Proposition 11.1]{Brylinski02b}. 

Assume now that $d=nr$ and let $\pi=[r^n]$ be the partition of $d$ 
into $n$ equal parts. Denote by $E_\pi$ the irreducible module of
the symmetric group $S_d$ which corresponds to $\pi$, and let $\chi$ 
be its character. Then by \cite[Corollary 11.1]{Brylinski02b} we have an
isomorphism
\begin{equation} \label{osnovna}
\pR_{n,k,d}^G \cong \left( E_\pi^{\otimes k} \right)^{S_d}
\end{equation}
of $S_k$-modules. The superscript $S_d$ means that one has to
form the space of invariants of $S_d$, i.e., the largest trivial
$S_d$ submodule of $E_\pi^{\otimes k}$. The other symmetric group,
$S_k$, acts on both sides by permuting the tensor factors.

This formula is very useful. For instance, one obtains immediately
the following formula for the dimension of the space of
$G$ invariants of degree $d$:
\begin{equation} \label{dim1}
\dim\, \pR_{n,k,d}^G =\frac{1}{d!} \sum_{g\in S_d} \chi(g)^k.
\end{equation}

By symmetrization, i.e., by taking the $S_k$-invariants on both 
sides of Eq. (\ref{osnovna}), and taking into account that the 
actions of $S_k$ and  $S_d$ commute, we obtain that
$$
\left( \pR_{n,k,d}^G \right)^{S_k} \cong 
\left( S^k( E_\pi ) \right)^{S_d}
$$
as complex vector spaces. (By $S^k(E_\pi)$ we denote the 
$k$-th degree piece of the symmetric algebra $S(E_\pi)$ of the
module $E_\pi$.)

By performing anti-symmetrization instead of the symmetrization,
one obtains a similar formula for the dimension of the space
of odd invariants of $S_k$ in $\pR_{n,k,d}^G$. Then on the
right hand side one should replace the symmetric power
$S^k(E_\pi)$ by the exterior power $\wedge^k(E_\pi)$.

The character $\chi^{(k)}$ of the $S_d$-module
$S^k(E_\pi)$ is given by the classical formula~\cite{Macdonald,FultonHarris}
\begin{equation} \label{kar-sim}
\chi^{(k)}(g)=\sum_{({\bf i})} \prod_{\alpha=1}^k
\frac{ \chi(g^\alpha)^{i_\alpha} }{ i_\alpha ! \alpha^{i_\alpha} },
\end{equation}
where the summation is over all sequences
$({\bf i})=(i_1,i_2,\ldots,i_k)$ of nonnegative integers such that
\[ \sum_\alpha \alpha\; i_\alpha = k. \]
This is valid for all permutations $g\in S_d$.

Similarly, the $S_d$-character $\chi^{[k]}$ of the $k$-th
exterior power $\wedge^k(E_\pi)$ is given by the formula
\begin{equation} \label{kar-kosi}
\chi^{[k]}(g)=\sum_{({\bf i})} \prod_{\alpha=1}^k 
\frac{ (-1)^{i_\alpha-1} \chi(g^\alpha)^{i_\alpha} }
{ i_\alpha ! \alpha^{i_\alpha} }.
\end{equation}

The values of the irreducible characters of $S_d$
are easily available, say in James and Kerber book~\cite{JamesKerber} 
or in software systems such as Maple or GAP.
Hence, we obtain the following formula for the space of joint
$G$ and $S_k$-invariants of degree $d=nr$:
$$
\dim\, \left( \pR_{n,k,d}^G \right)^{S_k} =
\frac{1}{d!}  \sum_{g\in S_d} \chi^{(k)}(g).
$$

In our case we have $n=2$, since we work with qubits, and $k=5$,
i.e., the number of qubits is 5. In that case there are exactly
seven sequences $({\bf i})$ having the required property.
Explicitly, they are:
$(5,0,0,0,0)$, $(3,1,0,0,0)$, $(1,2,0,0,0)$, $(2,0,1,0,0)$,
$(0,1,1,0,0)$, $(1,0,0,1,0)$ and $(0,0,0,0,1)$.
Formula~\eqref{kar-sim} then reads as
\begin{eqnarray*}
120\chi^{(5)}(g) &=& \chi(g)^5+10\chi(g)^3\chi(g^2)+15\chi(g)\chi(g^2)^2
+20\chi(g)^2\chi(g^3) \\
&& +20\chi(g^2)\chi(g^3)+30\chi(g)\chi(g^4)+24\chi(g^5).
\end{eqnarray*}

For instance, if $d=8=2\cdot4$ we have $r=4$, $\pi=[4,4]$, $E_\pi$ is the
module $X_8$, and
the values of $\chi$ on the representatives of the 22 conjugacy
classes of $S_8$ are 
\[ 14,4,2,0,6,-1,1,-1,2,-2,-2,0,2,1,2,-1,-1,-1,0,0,0,0 \]
(see James and Kerber, p. 351). By using the above formula, we find that
the values of the character $\chi^{(5)}$ on the same representatives are
\[ 8568,216,72,0,536,0,0,0,18,0,-12,0,12,0,24,3,1,0,0,2,0,0. \]
Then the multiplicity of the principal character (i.e., the
character of the trivial $S_8$-module $X_1$)
in $\chi^{(5)}$ is equal to the dimension of
the space of $SL^*$-invariants of degree 8.
Hence we have
\[ \dim {\rm Inv}_8^{SL^*} = \frac{1}{8!} \sum_{g\in S_8} \chi^{(5)}(g). \]
The evaluation of this sum confirms our finding that this dimension is $4$.

In conclusion, we summarize the results of our computations.
The number of linearly independent  $SL^*$-invariants in degrees 0, 2, 4,..., 22 is 
1, 0, 1, 0, 4, 0, 12, 2, 39, 21, 130, 115 respectively. 
The number of linearly independent relative $SL^*$-invariants in degrees
0, 2, 4,..., 22 is 1, 0, 1, 1, 4, 2, 14, 11, 49, 58, 185, 269 respectively.

\section{Connection between the $\Omega$-process and the comb approach}\label{altomega}

In this section we present a rephrasing of elements of Cayley's 
$\Omega$-process in terms of local invariant antilinear operators.
The central operations in the $\Omega$-process are 
determinants of derivatives
\beq\label{omega-def}
\Omega_x=\det\left|
\begin{array}{cc}
\partial_{x_0'} & \partial_{x_1'}\\
\partial_{x_0''} & \partial_{x_1''}
\end{array}\right|
\eeq
with subsequent ``trace'' $\trace: x',x''\rightarrow x$ applied to 
functions of the wave function 
coefficients $\psi_{i_1,\dots,i_q}$ dressed with auxiliary variables
$z^{(j)}_{i_j}$ such that a wave function
$\ket{\Psi}:=\sum\psi_{i_1,\dots,i_q} \ket{i_1,\dots,i_q}$ is mapped to the function
$f:=\sum \psi_{i_1,\dots,i_q}z^{(1)}_{i_1}\cdots z^{(q)}_{i_q}$.
A typical step in the $\Omega$-process is then prescribed as~\cite{Luque05}
$$
(P,Q)^{\epsilon_1,\dots,\epsilon_q}:=\trace
\Omega_{z^{(1)}}^{\epsilon_1}\cdots\Omega_{z^{(q)}}^{\epsilon_q}
P(z')Q(z'')\ .
$$
The key observation is that the action of $\Omega_x$~\eqref{omega-def}
amounts to a contraction of two of the
wave function coefficients with the antisymmetric tensor 
$\epsilon_{kl}$, $k,l\in\{0,1\}$
with $\epsilon_{01}:=1$ \footnote{It is worth 
emphasizing that $\epsilon=i\sigma_2$ has the physical interpretation of a 
{\em spinor-metric}~\cite{LandauLifshitzIII}.}. 
We illustrate this procedure in the most simple example
\nbeqa
B_{22200}&=&(f,f)^{0,0,0,1,1}\\
&=& \psi_{i_1,\dots,i_3,k,l}{\psi_{j_1,\dots,j_3}}^{k,l} 
z_{i_1}^{(1)} z_{i_2}^{(2)} z_{i_3}^{(3)} z_{j_1}^{(1)} z_{j_2}^{(2)} z_{j_3}^{(3)}, \\
\left. B_{22200}\right|_{z_.^{(.)}=1}&=&\psi_{i_1,\dots,i_3,k,l}{\psi_{j_1,\dots,j_3}}^{k,l}
=-(\!({\mathfrak I}{\mathfrak I}{\mathfrak I}\sigma_2\sigma_2)\!)\; ,
\neeqa
where we used Einstein sum convention and contraction via $\epsilon$.
The above so-called {\em transvectant} $B_{22200}$, which is bilinear in the 
$z^{(j)}$ ($j=1,2,3$), 
coincides with the subsequently shown antilinear expectation value
after setting all $z_{i_j}^{(j)}=1$;
here ${\mathfrak I}=\Matrix{cc}{1&1\\1&1}$.
It is seen that the pairs of
wave function copies to be contracted with $\epsilon$ 
by $\Omega_z$~\eqref{omega-def} 
are all those, whose functions $f$ contain the variable $z'$ 
and $z''$, respectively. Since both these variables might occur in more than 
one function $f$, the action of $\Omega_z$ will in general lead to a sum over
such $\epsilon$ contractions involving different pairs of copies of the 
wave function. 

Though it is clear now that each invariant constructed by the 
$\Omega$-process can be directly transcribed into a sum of complete 
contractions of the wave function coefficients via the antisymmetric 
tensor $\epsilon=i\sigma_2$,
it cannot be directly written in terms of antilinear expectation values
of $\sigma_2$.
A simple three qubit counterexample is the invariant 
\beq\label{omega-threetangle}
\tau_3=-2\psi_{a_1,a_2,a_3}{\psi^{a_1,a_2}}_{a_4}\;
       {\psi_{b_1,b_2}}^{a_3}\psi^{b_1,b_2,a_4}
      =(\!(\sigma_2\sigma_2\sigma_\mu 
       \bullet \sigma_2\sigma_2\sigma^\mu)\!)
\eeq
whose absolute value is the three-tangle~\cite{Coffman00}.
For obtaining the second equality in Eq.~\eqref{omega-threetangle}
we grouped in pairs the first and the last two wave function coefficients.
The contractions of the idexes $a_1$, $a_2$, $b_1$ and $b_2$ then appear 
inside these pairs and we will call them {\em inner contractions}.
On the other hand the contractions of the indexes $a_3$ and $a_4$ 
involve two different pairs of coefficients, and we will call them 
{\em cross-contractions}.
 
As stated above, for more complicated invariants produced by 
the $\Omega$-process, this correspondence is not given by a single 
full contraction; however, each of those complete contractions
contained in the $\Omega$-process is an invariant.
E.g. the invariant $F$ of degree $6$ (see Ref.~\cite{Luque05} for its 
construction from an $\Omega$-process) is 
equivalently expressed as
\begin{eqnarray} \label{F-epsilon}
F &=& 96 \, \psi_{i_1,i_2,i_3,i_4,i_5} \,
	\psi_{i_6,i_7}{}^{i_3}{}_{i_8}{}^{i_5} \,
	\psi_{i_9}{}^{i_2}{}_{i_{10}}{}^{i_4}{}_{i_{11}}
\\ \notag 
 &&	\psi^{i_1}{}_{i_2,i_3}{}^{i_8,i_{11}} \,
	\psi^{i_6,i_7,i_{13}}{}_{i_{14},i_{15}} \,
	\psi^{i_9,i_{12},i_{10},i_{14},i_{15}}.
\end{eqnarray}
Since the space of degree $6$ invariants for five qubits is one-dimensional,
the expression \eqref{F-epsilon} reproduces precisely this unique invariant $F$
(up to a prefactor). 
For the invariant of degree $10$ as constructed in Eq.~\eqref{omega10}, a 
possible transcription is
\begin{eqnarray} \label{G510-epsilon}
\tilde{\cal G}_{10}^{(5)} 
 &=&	\psi_{i_1,i_2,i_3,i_4,i_5} \,
	\psi_{i_6,i_7,i_8,i_9,i_{10}} \,
	\psi_{i_{11}}{}^{i_2,i_3,i_4}{}_{i_{12}} \,
	\psi^{i_6,i_7}{}_{i_{13},i_{14}}{}^{i_5}
\\ \notag 
 &&	\psi^{i_1}{}_{i_{15}}{}^{i_8}{}_{i_{16}}{}^{i_{10}} \,
	\psi_{i_{17}}{}^{i_{15}}{}_{i_{18}}{}^{i_9}{}_{i_{19}} \,
	\psi^{i_{11}}{}_{i_{20},i_{21}}{}^{i_{14},i_{12}}
\\ \notag 
 &&	\psi^{i_{22},i_{20},i_{13},i_{16},i_{25}} \,
	\psi_{i_{22},i_{23}}{}^{i_{18}}{}_{i_{24},i_{25}} \,
	\psi^{i_{17},i_{23},i_{21},i_{24},i_{19}}.
\end{eqnarray}
The space of degree $10$ invariants has dimension $15$, and the tilde 
indicates that the expression \eqref{G510-epsilon} cannot be expected 
to coincide with ${\cal G}^{(5)}_{10}$ as created from the $\Omega$-process. 
The latter is rather a sum
over all possible $\epsilon$-contractions emerging from the given 
$\Omega$-process, and 
$\widetilde{\cal G}^{(5)}_{10}$ is only one element of this sum.
This transcription therefore already reduces the computational 
complexity of such invariants.
Interestingly, the symmetric group $S_5$ generates from 
$\widetilde{\cal G}^{(5)}_{10}$ a $14$-dimensional subspace
where only $PF$ is missing to give the whole $15$-dimensional
space $V^{(5)}_{10}$.

\begin{figure}[h]
\includegraphics[width=.8\textwidth]{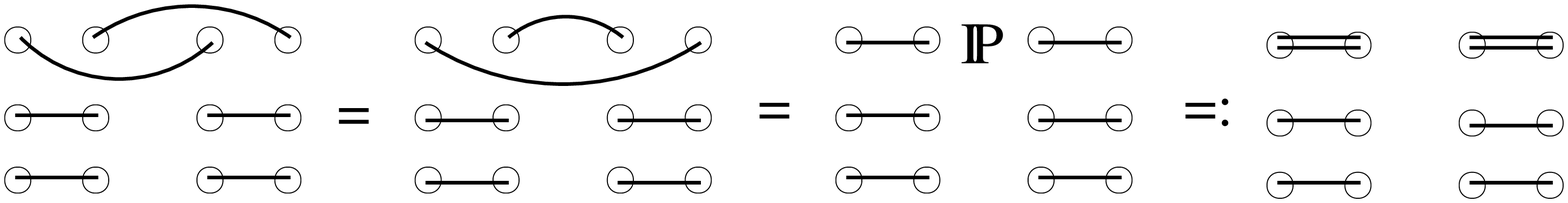}
\caption{Three-qubit wavefunction coefficients
sketched as a staple of three circles. Each contraction 
with the antisymmetric tensor $\epsilon=i\sigma_2$ is visualized by a line 
connecting two circles. Arranged in pairs,
an expectation value with $\epsilon$ corresponds to an inner contraction.
\label{IdentI}}
\end{figure}

A view back onto Eq.~\eqref{omega-threetangle} 
suggests a connection between the $\Omega$-process and 
the invariant construction from combs; namely that the cross-contraction on the third qubit 
might be substituted by the comb of second order $\sigma_\mu\bullet\sigma^\mu$,
possibly including a term proportional to $\sigma_2\bullet\sigma_2$.
In order to make this connection a rigorous statement we translate 
the index contraction into an antilinear expectation value.
The symmetry of antilinear expectation values
\beq\label{alin-symm}
(\psi^*|\hat{A}|\varphi)=(\varphi^*|\hat{A}^\dagger|\psi)
=(\varphi^*|\hat{A}|\psi)
\eeq
for Hermitean operators is crucial for this to work. 
The procedure is best explained graphically in figure~\ref{IdentI}.
The three-qubit wavefunction coefficients are
sketched as a staple of three circles there. They are arranged in pairs,
and an expectation value corresponds to an inner contraction of that pair
-- i.e. a contraction of wavefunction indexes inside such a pair.
Each contraction with the antisymmetric tensor $\epsilon=i\sigma_2$ 
is visualized by a line connecting two circles. 
The cross-contractions are not yet expressed in terms of expectation values.
Fortunately, suitable permutations of copies, which are however local 
in the qubits, do exist as to transform also the cross-contractions 
into expectation values without disturbing the inner contractions.
In the following we describe this iterative procedure.
The first equality in Fig.~\ref{IdentI} is due to the symmetry
\eqref{alin-symm}
for antilinear expectation values of Hermitean operators.
The second equality is formally expressed as
\nbeqa
\bra{\psi^*}\bullet\bra{\psi^*}\mathop{\P}^3\sigma_2\sigma_2\sigma_2\bullet\sigma_2\sigma_2\sigma_2 
\ket{\psi}\bullet\ket{\psi} &=&\bra{\psi^*}\bullet\bra{
\psi^*}\sigma_2\sigma_2\sigma_2\bullet\sigma_2\sigma_2\sigma_2 \mathop{\P}^3
\ket{\psi}\bullet\ket{\psi}\\ 
&=& -\psi_{a_1\,a_2\,a_3}{\psi^{a_1\,a_2}}_{a_4}
                               {\psi_{b_1\,b_2}}^{a_3}\psi^{b_1\,b_2\,a_4}
\neeqa
where $\P$ is the symbol for a copy permutation operator 
and the number three on top of $\mathop{\P}$ indicates that this 
permutation operator acts non-trivially only on the third qubit. 
Using $\P=\frac{1}{2}\sum_{\mu=0}^3 \sigma_\mu\bullet\sigma_\mu$, 
a straightforward calculation produces
\beqb{rcl}{Omega-Pi}
(\sigma_2\bullet\sigma_2)\,\P&=&M_{\mu\nu}\sigma_\mu\bullet\sigma_\nu\\
&=&-\frac{1}{2}\left(\sigma_\mu\bullet\sigma^\mu - \sigma_2\bullet\sigma_2\right)
\eeqb
where $M_{\mu\nu}=\delta_{\mu\nu}m_\mu$, $(m_0,m_1,m_2,m_3)=(1,-1,1,-1)/2$.
The resulting antilinear expectation value of 
$M_{\mu\nu}\sigma_\mu\bullet\sigma_\nu$
is then indicated graphically by a double line connecting the copies.
For completeness we mention that 
$$
(\sigma_\mu\bullet\sigma^\mu)\,\P=
-\frac{1}{2}\left(\sigma_\mu\bullet\sigma^\mu +3 \sigma_2\bullet\sigma_2\right)
$$
which readily follows from the identity 
$(\sigma_2\bullet\sigma_2)\,\P \P=(\sigma_2\bullet\sigma_2)$ and 
Eq.~\eqref{Omega-Pi}.

\begin{figure}[h]
\includegraphics[width=.8\textwidth]{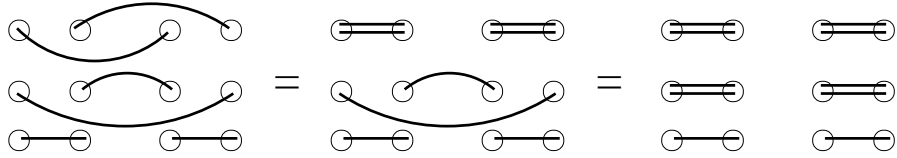}
\caption{Using copy permutation operators that act separately on demand on
specific qubits, the above graphical expression for an identity is obtained.
It shows that the rephrasing in terms of antilinear expectation values
can indeed be obtained in this way.}\label{SecondIdentity}
\end{figure}

In order to see in how far every invariant constructed
with an $\Omega$-process can be expressed in terms of expectation values 
of local invariant operators, the identity described in Fig.~\ref{SecondIdentity} 
is helpful.

It means that the translation of the $\Omega$-process 
into antilinear expectation values can be performed iteratively qubit per
qubit: an apparent incompatibility of a contraction pattern with a 
fixed ordering of wave function coefficients (vertical rows of circles) 
is resolved iteratively making use of the symmetry \eqref{alin-symm}.
It is worth mentioning at this point that grouping the wavefunction 
coefficients in pairs is arbitrary and therefore we obtain
the same invariant when changing this order by permuting the
order of the wave function coefficients. This however changes the
contraction scheme for the invariant and leads to identities
for invariants as observed above. For degree $4$ invariants this freedom 
means that interchanging 
$$
\sigma_2\bullet\sigma_2\leftrightarrow 
\frac{1}{2}(\sigma_2\bullet\sigma_2- \sigma_\mu\bullet\sigma^\mu)
$$
on each qubit leads to the same invariant. This immediately leads to
the identities~\eqref{mmmmId}, those mentioned just above it,
and the identities~\eqref{mmm22Id} and~\eqref{PsymmId}.

For higher degree invariants, more such pair permutations may occur. 
Since the symmetric group is generated from 
nearest neighbor exchanges $\pi_{j,j+1}$ and by virtue of the relations
$\pi_{12}\pi_{23}\pi_{12}=\pi_{23}\pi_{12}\pi_{23}$, $\pi_{ij}^2=\id$,
it is sufficient to consider the results for up to three permutation operators.
We find that
\beqb{ccl}{Omega-PiPi}
(\sigma_2\bullet\sigma_2\bullet\sigma_2)\,\P_{12}\P_{23}
&=& \frac{1}{4}\left[\sigma_2\bullet\sigma_2\bullet\sigma_2-\right.\\
&& \left(\sigma_\mu\bullet\sigma^\mu\bullet\sigma_2 
  + \sigma_\mu\bullet\sigma_2\bullet\sigma^\mu 
  + \sigma_2\bullet\sigma_\mu\bullet\sigma^\mu \right)\\
&& \left. - i \epsilon_{klm} \tau_k\bullet\tau_l\bullet\tau_m\right]
\eeqb
and
\beqb{ccl}{Omega-PiPi2}
(\sigma_2\bullet\sigma_2\bullet\sigma_2)\,\P_{23}\P_{12}
&=& \frac{1}{4}\left[\sigma_2\bullet\sigma_2\bullet\sigma_2-\right.\\
&& \left(\sigma_\mu\bullet\sigma^\mu\bullet\sigma_2 
  + \sigma_\mu\bullet\sigma_2\bullet\sigma^\mu 
  + \sigma_2\bullet\sigma_\mu\bullet\sigma^\mu \right)\\
&& \left. + i \epsilon_{klm} \tau_k\bullet\tau_l\bullet\tau_m\right]\ ,
\eeqb
where $\tau_1:=\sigma_0$, $\tau_2:=\sigma_1$, $\tau_3:=\sigma_3$. 
A further permutation acting on the antisymmetric part 
$i \epsilon_{klm} \tau_k\bullet\tau_l\bullet\tau_m$ leads to terms of the type 
$\left(\sigma_\mu\bullet\sigma^\mu\bullet\sigma_2 
  + \sigma_\mu\bullet\sigma_2\bullet\sigma^\mu 
  + \sigma_2\bullet\sigma_\mu\bullet\sigma^\mu \right)$ and vice versa.

Besides an antisymmetric part in the exchange of copies,
only $\sigma_2$ and $\sigma_\mu\bullet\sigma^\mu$ appear and the
set of locally invariant operators is closed.
The antisymmetric part is not captured by the
two combs but in our analysis it appeared to be irrelevant for the search of
$SL^*$-invariants. It leads, however, to invariants 
that are antisymmetric under qubit permutations
(see e.g. the invariants of degree $6$ and $10$ for five qubits).
Since an entanglement measure is defined as the modulus of an invariant,
the requirement for a class-specific global entanglement measure 
must be relaxed to include also $SL^*_-$-invariants.

\section{Conclusions}\label{concls} 

We have presented a thorough analysis of the polynomial $SL$-invariants of 
four and five qubits with particular emphasis on the filter ideal, i.e. on
those invariants that vanish on all product states. 
Within the complete ring of invariants the filter ideal plays an 
outstanding role because only these invariants can clearly discriminate 
the genuine multipartite entanglement from that within parts of the system.
It therefore hosts candidates for a class specific 
quantification of genuine multipartite entanglement
and its knowledge plays a key role for a systematic analysis and a deeper 
understanding of the structure of entanglement in multipartite systems.
A reasonable measure for global entanglement should also be invariant under
qubit permutations. 
In order to determine the dimension of the subspace of permutation 
invariant elements in the filter ideal, we analyzed the decomposition 
of the space of polynomial invariants into irreducible 
modules of the symmetric group. 

In the case of four qubits, the standard approach from invariant theory, 
employing the well established $\Omega$-process by Cayley, 
has already lead to the construction of
a complete set of $SL$-invariants~\cite{Luque05}. 
We have compared this approach to an alternative proposal based on 
local invariant operators, termed combs~\cite{OS04,OS05}. 
We could demonstrate that also the latter approach generates a complete set of 
invariants, and we provide a full dictionary for expressions from 
both approaches.
We have furthermore established the equivalence of the $\Omega$-process and the 
contraction with the spinor metric and provide the missing 
link between the $\Omega$-process and the construction from local 
$SL$-invariant operators. This implies that major part of the ring of invariants
can be generated using the comb based method. Indeed we find that all 
$SL^{*}$-invariants and even many $SL^{*}_-$-invariants in this work are 
accessible to the comb based approach.
In addition, the interrelation between $\Omega$-process and the comb approach
implies the existence of interesting identities among sets of invariants and
readily explains those identities observed among invariants of degree $4$. 

We single out two major advantages of the approach from local invariant
operators.\\
\begin{itemize}
\item[I]
{\bf Control}:
The comb based approach admits a high degree of control over specific 
important properties of the invariants that are to be constructed. 
Of particular relevance is the ab initio knowledge about the set of 
product states for which the invariant will vanish.
This is a key quality that admits a systematic construction of filter 
invariants; this provides a targeted construction of the filter ideal.
In contrast, from the $\Omega$-process, and equivalently from the contraction 
with the spinor metric, there is no a priori knowledge about the 
invariant's value on product states.
\item[II]
{\bf Computational complexity}:
The interrelation between $\Omega$-process, $\epsilon$ contractions and
the comb based approach explains the notable difference in 
computational complexity we observed. 
It is clear from these interrelations that the computational complexity 
of the comb based approach is significantly lower than that using contractions 
with the spinor metric $\epsilon$, which itself already constitutes a 
speed-up as compared to the $\Omega$-process. 
This discrepancy grows more important with increasing 
degree of the invariants.
\end{itemize}
These advantages permit us to go considerably further in a 
thorough analysis of invariants, and we demonstrate this
for five qubits: we give a complete analysis of $SL$-invariants up to 
degree $12$, and provide an outlook on the situation for degrees $14$ and $16$. 
Although the five qubit case is still far from being completed, we have presented
a straightforward technique for how to proceed; we are confident that a 
minimal set of generators can be obtained in the way described in this manuscript.
All results have been cross-checked with predictions from the Hilbert series. 
To this end we also derived the first terms of the Hilbert series for 
relative $SL^{*}$-invariants.

We hope that the high degree of control paired with the significantly
lower computational complexity in the generation of invariants 
have future impact in both quantum information and invariant theory. 
Further analysis would be necessary in order to find an expression of 
all $SL^{*}_-$-invariants in terms of (antilinear) expectation values.
It would be also interesting to extend an analysis along the lines proposed
e.g. in Refs.~\cite{Levay06,Levay05} in order
to see whether the filter ideal has a distinguished 
geometrical interpretation.

\begin{appendix}

\section{Comb-based invariants}
\label{combs}
In this appendix we give a detailed elucidation how comb-based invariants
are calculated.

Let the pure $q$-qubit quantum state $\ket{\psi}$ be expressed in terms
of a basis ${\cal B}$ 
made of tensor products of eigenstates $\ket{-1}$
and $\ket{1}$ of the Pauli spin operator $\sigma_3$,
such that $\sigma_3 \ket{s}= s \ket{s}$ for $s=\pm 1$.
That is, we have 
$$
{\cal B} = \left\{\, \ket{s_1}\otimes\cdots\otimes\ket{s_q}\; :\; 
s_j=\pm 1\, \right\}
$$
In this basis the Pauli spin operators (consider $q=1$ for the sake of simplicity)
assume the matrix representations
$\sigma_i^{s's}:=\bra{s'}\sigma_i\ket{s}$ as given 
in Eq.~\eqref{Pauli}.
Matrix elements of $q$-qubit operators are then defined in the standard way
for arbitrary $q$-qubit pure states $\ket{\varphi}$, $\ket{\psi}$ as
\beqa\label{MatElms}
\bra{\varphi}\sigma_{i_1}\otimes\cdots\otimes\sigma_{i_q}\ket{\psi}&:=&
(\varphi_{s'_1,\dots,s'_q})^*
\sigma_{i_1}^{s'_1,s_1}\cdots\sigma_{i_q}^{s'_q,s_q}
\psi_{s_1,\dots,s_q}\\
&\equiv&
[{\mathfrak C}\varphi]_{s'_1,\dots,s'_q}
\sigma_{i_1}^{s'_1,s_1}\cdots\sigma_{i_q}^{s'_q,s_q}
\psi_{s_1,\dots,s_q}\label{Conjug}
\eeqa
within Einstein summation convention, and 
$[{\mathfrak C}\varphi]_{s_1,\dots,s_q}:=
\bra{s_1,\dots,s_q}{\mathfrak C}\ket{\varphi}$.

In this sense the antilinear expectation values as defined in 
Eq.~\eqref{doubleparenthesis}
are specific matrix elements of an antilinear operator $A=L_A{\mathfrak C}$.
Here, $\mathfrak C$ is the complex conjugation 
as defined in Eq.~\eqref{Conjug},
and $L_A$ is the linear operator associated to $A$.
In all this work, the operators $A$ and $L_A$ are antilinear Hermitean
and Hermitean, respectively. 
In the case of a single copy of the state
we then have
$$
(\!(L_A)\!):=\bra{\psi}A^\dagger\ket{\psi}^*
=\bra{\psi}A\ket{\psi}^*
=\bra{\psi^*}L_A\ket{\psi}
$$
which is a matrix element as defined in Eq.\eqref{MatElms}
where $\ket{\varphi}\rightarrow \ket{\psi^*}$ (see Eq.~\eqref{Conjug}).
For any indexes $i_1,i_2,\ldots,i_q\in\{0,1,2,3\}$ we therefore define a
bilinear form
$$
\la \sig_{i_1} \sig_{i_2} \cdots \sig_{i_q} \ra : \pH_q \times \pH_q\to\CC,
$$
whose value at $(\vf,\psi)$ is the multiple sum (using the Einstein convention)
$$ \sig_{i_1}^{a_1,b_1} \sig_{i_2}^{a_2,b_2} \cdots \sig_{i_q}^{a_q,b_q}
\vf_{a_1,a_2,\ldots,a_q} \psi_{b_1,b_2,\ldots,b_q}. $$

This can be also expressed as
$$
\la \sig_{i_1} \sig_{i_2} \cdots \sig_{i_q} \ra (\vf,\psi)
= \la \vf^* | \sig_{i_1} \otimes \sig_{i_2} \otimes \cdots \otimes \sig_{i_q} | \psi \ra .
$$

As the first example, we set $q=1$ and $i_1=2$ and we obtain the 
$SL(2,\CC)$-invariant bilinear form
\beq\label{1-comb}
 \la \sig_2 \ra ( \vf,\psi )=-i\left|
\begin{array}{ll} \vf_1 & \vf_2 \\ \psi_1 & \psi_2 \end{array} \right|. 
\eeq
However, in this case we have 
$(\!(\sig_2)\!):= \la \sig_2 \ra ( \psi,\psi )=0$
for all $\psi$,
which is the comb property of the operator $\sigma_2$.

As another example we take $q=2$ and $i_1=i_2=2$. Since
$\sig_2=-i\epsilon$, we have
\begin{eqnarray*}
\la \sig_2 \sig_2 \ra ( \vf,\psi ) &=& -\epsilon^{a_1,b_1} \epsilon^{a_2,b_2}
\vf_{a_1,a_2} \psi_{b_1,b_2} \\
&=& \left| \begin{array}{ll} \vf_{2,1} & \vf_{2,2} \\ 
\psi_{1,1} & \psi_{1,2} \end{array} \right|-
\left| \begin{array}{ll} \vf_{1,1} & \vf_{1,2} \\ 
\psi_{2,1} & \psi_{2,2} \end{array} \right|,
\end{eqnarray*}
an $SL$-invariant bilinear form.
In the case when $\vf=\psi$, we obtain the nonzero $SL$-invariant quadratic form
$$(\!(\sig_2 \sig_2)\!):= \la \sig_2 \sig_2 \ra ( \psi,\psi ) = -2 \left| \begin{array}{ll}
 \psi_{1,1} & \psi_{1,2} \\ 
\psi_{2,1} & \psi_{2,2} \end{array} \right|. $$

For operators acting on $m$ copies of the state just replace
$\ket{\psi}$ by 
$\ket{\psi}\bullet\dots\bullet\ket{\psi}=:\ket{\psi}^{\bullet m}$
and the corresponding expression for $\bra{\psi^*}$.
To outline this in more detail, 
let $\pH$ denote the Hilbert space of a single qubit,
and $\pH_q=\pH^{\otimes q}$ the one for the system of $q$ qubits.
We shall also use the Hilbert space for $m$ copies of this multipartite 
system. In that case we use $\bullet$ to denote tensor products of
Hilbert spaces of different copies.
Let us now take a collection of $m$ bilinear forms of the above type,
$$ \la \sig_{{i_1}^{(k)}} \sig_{{i_2}^{(k)}} \cdots \sig_{{i_q}^{(k)}} \ra ,
\quad k=1,2,\ldots,m, $$
and let us form their product
\begin{equation} \label{TenzFor}
\la \sig_{{i_1}^{(1)}} \sig_{{i_2}^{(1)}} \cdots \sig_{{i_q}^{(1)}} \ra
\la \sig_{{i_1}^{(2)}} \sig_{{i_2}^{(2)}} \cdots \sig_{{i_q}^{(2)}} \ra
\cdots
\la \sig_{{i_1}^{(m)}} \sig_{{i_2}^{(m)}} \cdots \sig_{{i_q}^{(m)}} \ra,
\end{equation}
which is a bilinear form
\begin{equation} \label{GenFor}
\pH_q^{\bullet m} \times \pH_q^{\bullet m} \to \CC .
\end{equation}
The value of this bilinear form on the special elements
$$ ( \vf^{(1)} \bullet \vf^{(2)} \bullet \cdots \bullet \vf^{(m)},
 \psi^{(1)} \bullet \psi^{(2)} \bullet \cdots \bullet \psi^{(m)} )
$$ 
is equal to
$$
\prod_{k=1}^{m} \la \sig_{{i_1}^{(k)}} \sig_{{i_2}^{(k)}} 
\cdots \sig_{{i_q}^{(k)}} \ra \left( \vf^{(k)}, \psi^{(k)} \right).
$$
If we further specialize $ \vf^{(k)} = \psi^{(k)} = \psi $
for all $k$, we obtain the $2^q$-ary form of degree $2m$ in the complex
components of $\psi$:
$$
(\!(\prod_{k=1}^{\bullet m} \sig_{{i_1}^{(k)}}  
\cdots \sig_{{i_q}^{(k)}})\!):=
\prod_{k=1}^m \la \sig_{{i_1}^{(k)}}
\cdots \sig_{{i_q}^{(k)}} \ra \left( \psi,\psi \right).
$$
We refer to this form of degree $2m$ as the {\em associated form}
of the bilinear form (\ref{TenzFor}). This definition extends 
immediately to any bilinear form (\ref{GenFor}).

In general, the forms of degree $2m$ constructed above are not 
$SL$-invariant, but we can use their
suitable linear combinations to obtain $SL$-invariant forms.

In order to do that we proceed as follows. First we select a site,
say $s$, $1\le s\le q$, of our multipartite system and 
choose two different copies of the state, say copies $p$ and $q$,
$1\le p<q\le m$. Next we replace in (\ref{TenzFor}) the Pauli matrices
$\sig_{{i_s}^{(p)}}$ and $\sig_{{i_s}^{(q)}}$ with symbols
$\sig_\mu$ and $\sig^\mu$, respectively. This is to indicate that
the two idexes $\mu$ are to be contracted by using the
pseudo-metric $G_{\mu\nu}$. We now interrupt our description to give
an example.

When $q=3$ and $m=2$ the expression (\ref{TenzFor}) has the form
$$
\la \sig_{{i_1}^{(1)}} \sig_{{i_2}^{(1)}} \sig_{{i_3}^{(1)}} \ra
\la \sig_{{i_1}^{(2)}} \sig_{{i_2}^{(2)}} \sig_{{i_3}^{(2)}} \ra
$$
We now choose $p=1$, $q=2$ and $s=1$. By applying the above instruction,
we obtain the expression
$$
\la \sig_\mu \sig_{{i_2}^{(1)}} \sig_{{i_3}^{(1)}} \ra
\la \sig^\mu \sig_{{i_2}^{(2)}} \sig_{{i_3}^{(2)}} \ra\; .
$$
By fixing $\sig_{{i_2}^{(1)}}=\sig_{{i_3}^{(1)}}=\sig_{{i_2}^{(2)}}=
\sig_{{i_3}^{(2)}}=\sig_2$ and performing the $\mu$-contraction this
gives the linear combination
$$
-\la \sig_0 \sig_2 \sig_2 \ra \la \sig_0 \sig_2 \sig_2 \ra
+\la \sig_1 \sig_2 \sig_2 \ra \la \sig_1 \sig_2 \sig_2 \ra
+\la \sig_3 \sig_2 \sig_2 \ra \la \sig_3 \sig_2 \sig_2 \ra .
$$
The associated quartic form is then obtained as
\nbeqa
(\!(\sigma_\mu \sigma_2 \sigma_2
&\bullet&\sigma^\mu \sigma_2 \sigma_2)\!):= 
\sum_{\mu=0}^3 g_\mu \bra{\psi^*}\sigma_\mu \sigma_2 \sigma_2\ket{\psi}^2\\
&=& \sum_{\mu=0}^3 g_\mu \left[\psi_{s'_1,s'_2,s'_3}
\sigma_{\mu}^{s'_1,s_1}\sigma_{2}^{s'_2,s_2}\sigma_{2}^{s'_3,s_3}
\psi_{s_1,s_2,s_3}\right]^2\; .
\neeqa
It generates the $SL^*$-invariants for three qubits; 
its modulus is the three-tangle~\cite{Coffman00}.

To continue our description, we choose a collection of triples
$(s_i,p_i,q_i)$, $i=1,2,\ldots,t$ such that $1\le p_i<q_i\le m$ and
whenever $s_i=s_j$, with $i\ne j$, we require that the four integers 
$p_i,q_i,p_j,q_j$ be all distinct. For each index $i$, we replace 
the Pauli matrices on the site $s_i$ and copies $p_i$ and $q_i$ 
with the symbols $\sig_{\mu_i}$ and $\sig^{\mu_i}$, respectively.
Next we replace all other Pauli matrices in (\ref{TenzFor})
with the matrix $\sig_2$. Finally, by using the pseudo-metric 
$G_{\mu\nu}$, we perform the $\mu_i$ contractions for each $i$,
$1\le i\le t$. We obtain a linear combination of bilinear forms
of the type given by (\ref{TenzFor}). We refer to these linear 
combinations as {\em comb-based bilinear forms}.

These comb-based forms are homogeneous multilinear expressions in the
(complex) state coefficients, which are $SL_q$-invariant. This invariance 
harkens back to the $SL(2,\CC)$ invariance of the antilinear single 
qubit combs $\sigma_2\mathfrak C$ and 
$\sigma_\mu{\mathfrak C}\bullet\sigma^\mu {\mathfrak C}$.
We formulate this statement in
\begin{theorem} \label{basic}
Any comb-based bilinear form (and, consequently, also its
associated form) is an $SL$-invariant. 
\end{theorem}

It has been stated in~\cite{OS04,OS05} that the combs are $SL$-invariant, 
but there is only implicit reference to the fact that this derives from
the central comb property to have zero expectation value on all the local 
Hilbert spaces. Here we sketch a proof for this connection.

{\em Proof: } The comb property for the operator $\sigma_2$, namely that 
$\la \sigma_2 \ra (\psi,\psi)=0$ for all single qubit states $\psi$, 
can be read off directly from Eq.~\eqref{1-comb},
and it can be checked by direct calculation
that it is the unique operator with this property up to rescaling.
Also by direct calculation we find that 
\mbox{$\la \sigma_\mu\bullet\sigma^\mu \ra (\psi\bullet\psi,\psi\bullet \psi)=0$} 
for all single qubit states $\psi$.
Furthermore, this is the unique form (up to rescaling)
on $\pH_q^{\bullet m}$ satisfying 
this condition which is symmetric under copy-permutation and orthogonal
to $\la \sigma_2\bullet\sigma_2 \ra$ in the sense of the vanishing 
scalar product
$\trace (\sigma_2\bullet\sigma_2)\cdot(\sigma_\mu\bullet\sigma^\mu)=0$.
For arbitrary $S\in SL(2,\CC)$ we then find that
$$
0= \la \sigma_2 \ra (S\psi,S\psi)=\la S^t\sigma_2 S \ra (\psi,\psi)
$$
for all $\psi$ and, due to the uniqueness property for the operator $\sigma_2$,
this implies $S^t\sigma_2 S=\sigma_2$. Analogously we have
\nbeqa
0 &=& \la \sigma_\mu\bullet\sigma^\mu \ra (S\psi\bullet S\psi,S\psi\bullet S\psi)\\
&=& \la (S\bullet S)^t\sigma_\mu\bullet\sigma^\mu(S\bullet S) \ra 
(\psi\bullet \psi,\psi\bullet \psi)
\neeqa
This proves that the two comb operators are $SL(2,\CC)$-invariant. Consequently, 
a $q$-qubit form constructed from those is seen to be $SL_q$-invariant 
by wrapping a transformation $S^{(q)}=S_1\otimes\dots\otimes S_q$ 
with $S_j\in SL(2,\CC)$ back onto the states.
This completes the proof.

We refer to the $SL$-invariants constructed in this manner
as the {\em comb-based invariants}. In some cases this 
invariant may be zero.

\end{appendix}

\begin{theacknowledgments}
We acknowledge discussions with Jens Siewert and Mary Beth Ruskai for 
useful suggestions on the manuscript. One of the authors (D.D.) 
was supported in part by an NSERC Discovery Grant.
\end{theacknowledgments}

\end{document}